\documentclass[default,iicol]{sn-jnl}

\usepackage{graphicx}
\usepackage{amssymb}
\usepackage{float}
\usepackage{color}
\usepackage{times}
\usepackage{amsmath}
\usepackage{xr}
\newcommand{\AF}{A_{\rm FWM}}
\newcommand{\PhiF}{\Phi_{\rm FWM}}

\newcommand{\Ar}{A_{\rm 2r}}
\newcommand{\Phir}{\Phi_{\rm 2r}}

\unnumbered

\begin{document}

\title{Correlative light electron microscopy using small gold nanoparticles as single probes}


\author[1]{\fnm{Iestyn} \sur{Pope}}

\author[2,4]{\fnm{Hugh} \sur{Tanner}}

\author[1]{\fnm{Francesco} \sur{Masia}}

\author[1]{\fnm{Lukas} \sur{Payne}}

\author[2,5]{\fnm{Kenton Paul} \sur{Arkill}}

\author[2]{\fnm{Judith} \sur{Mantell}}

\author[3]{\fnm{Wolfgang} \sur{Langbein}}

\author*[1]{\fnm{Paola} \sur{Borri}}\email{BorriP@cardiff.ac.uk}

\author*[2]{\fnm{Paul} \sur{Verkade}}\email{P.Verkade@bristol.ac.uk}

\affil[1]{\orgdiv{School of Biosciences}, \orgname{Cardiff University}, \orgaddress{\street{Museum Avenue}, \city{Cardiff}, \postcode{CF10 3AX}, \country{UK}}}

\affil[2]{\orgdiv{School of Biochemistry}, \orgname{University of Bristol}, \orgaddress{\street{University Walk}, \city{Bristol}, \country{UK}}}

\affil[3]{\orgdiv{School of Physics and Astronomy}, \orgname{Cardiff University}, \orgaddress{\street{The Parade}, \city{Cardiff}, \postcode{CF24 3AA}, \country{UK}}}

\affil[4]{Present Address: \orgdiv{Department of Chemistry}, \orgname{Ume\r{a} University}, \city{Ume\r{a}}, \postcode{90187}, \country{Sweden}}

\affil[5]{Present Address: \orgdiv{School of Medicine}, \orgname{University of Nottingham}, \city{Nottingham}, \postcode{NG7 2RD}, \country{UK}}


\abstract{Correlative light electron microscopy (CLEM) requires the availability of robust probes which are visible both in light and electron microscopy. Here we demonstrate a CLEM approach using small gold nanoparticles as a single probe. Individual gold nanoparticles bound to the epidermal growth factor protein were located with nanometric precision background-free in human cancer cells by light microscopy using resonant four-wave-mixing (FWM), and were correlatively mapped with high accuracy to the corresponding transmission electron microscopy images. We used nanoparticles of 10\,nm and 5\,nm radius, and show a correlation accuracy below 60\,nm over an area larger than 10\,\textmu m size, without the need for additional fiducial markers. Correlation accuracy was improved to below 40\,nm by reducing systematic errors, while the localisation precision is below 10\,nm. Polarisation-resolved FWM correlates with nanoparticle shapes, promising for multiplexing by shape recognition in future applications. Owing to the photostability of gold nanoparticles and the applicability of FWM microscopy to living cells, FWM-CLEM opens up a powerful alternative to fluorescence-based methods.}




\maketitle
\thispagestyle{empty}
\pagestyle{empty}

\section{Introduction}\label{sec1}

Correlative light electron microscopy (CLEM) combines the strengths of light microscopy (LM) and electron microscopy (EM) and is receiving growing attention in the life sciences, especially after the recent revolutionary developments of super-resolution (SR) light microscopy and cryo-EM\,\cite{AndoJPDAP18,BookVerkade21}. CLEM aims to combine the live cell imaging capability, large field of views, and molecular specificity of LM with the spatial resolution and ultrastructural information of EM, to pin-point specific events and visualise molecular components in the context of the underlying intracellular structure at nanometric to atomic resolution. To highlight biomolecules of interest and determine their position with high accuracy in this context, they need to be labelled with probes that are visible both in the light microscope (typically by fluorescence) and in the electron microscope (electron dense material). The production and detection of appropriate probes for each imaging modality is one of the key aspects in any correlative microscopy workflow. 

A commonly used approach is to combine a fluorescent moiety together with a gold nanoparticle (AuNP)\,\cite{BrownProtoplasma2010,TannerMCB21}. Such dual probes can be made fairly easily and are also available commercially. For example, we have used an Alexa594 fluorescent dye and a 5\,nm diameter AuNP coupled to the ligand transferrin (Tf), a molecule that normally recycles between the plasma membrane and early endosomes. Importantly, we showed that such a conjugate was trafficking as expected, i.e. the function of Tf was not perturbed by the probe\,\cite{BrownProtoplasma2010}. However, the fluorescence of Tf-Alexa594 with the AuNP was diminished compared to Tf-Alexa594. Indeed, fluorescence quenching, due to nonradiative transfer in the vicinity of a AuNP, is a well documented effect, which can significantly reduce the applicability of these probes in CLEM workflows \cite{KandelaScanning07,MilesSR17}. Moreover, we have shown recently that the integrity of this type of dual probes inside cells, and in turn their ability to correlatively report the location of the same molecule, should be seriously questioned\,\cite{GiannakopoulouNS20}. 

Ideally one would like to use a single probe that is visible both in the light and in the electron microscope. Semiconductor nanocrystals, also called quantum dots (QDs), do represent a single CLEM probe as they harbour an electron dense core that also emits fluorescence\,\cite{GiepmansNM05}. However, QDs typically contain cyto-toxic atoms (e.g. Cd or As). In turn, they require a protective shell coating for bio-applications which can double the probe size\,\cite{GiepmansNM05}. Moreover, QDs have an intermittent ‘on-off’ emission (i.e. they blink)\,\cite{YuangACSNano18}. This limits their application e.g. in time-course experiments aimed at tracking the same probe over time, whereby blinking causes problems when trying to reconnect positions to generate long trajectories.   

Alternately, there have been some developments toward using fluorophores as single probes\,\cite{KukulskiJCB11,JohnsonSR15}. However, this is challenging since the fixation and staining protocols for EM are often not compatible with retaining fluorescence emission. Fluorescence imaging after sample preparation for EM is key to minimise the uncertainty regarding the relative positions of fluorescent labels and EM structural features, due to the anisotropic shrinking and deformations
caused by the sample processing steps. With the advent of cryo-EM which can directly image biomaterials without staining and offers the best approach to preserve the native cellular ultrastructure, workflows have been developed to perform light microscopy at cryogenic temperature\,\cite{SartoriJSB07,TuijtelSC19}. Notably, cryo-LM has the added benefit of an increased photostability of organic fluorophores at low temperature, which has been exploited to achieve super-resolution fluorescence microscopy\cite{HoffmanScience20,TuijtelSC19}, reducing the resolution gap between LM and EM modalities. However, cryo-LM is technically challenging, often requiring sophisticated custom setups with highly stable cryostages, and specific high NA long-working-distance air objectives to avoid sample devitrification. Moreover, the requirement for high light intensities onto the sample to achieve SR can cause sample devitrification and damage, and preclude subsequent imaging using cryo-EM. It is also important to point out that to achieve the highest correlation accuracy between LM and EM images, the addition of spherical bead fiducial markers that are visible in both modalities is typically required\,\cite{KukulskiJCB11,TuijtelSC19}. By measuring and matching the coordinates of the centroid of each fiducial marker in the LM image and the EM image, one can calculate the transformation between the two images, which takes into account changes in magnification, rotation, and distortions. However, introducing fiducials adds further steps to the sample preparation protocols, increasing complexity and possible artefacts by induced modifications. 

Another approach would be to use small AuNPs as single probes. These are easily visible in EM, and exhibit strong light scattering and absorption at their localised surface plasmon resonance (LSPR). They are photostable, and the achievable photon fluxes are governed by the incident photon fluxes and the AuNP optical extinction cross-section, a significant advantage compared to fluorophores which can emit a maximum of one photon per excited-state lifetime. However, when small AuNPs are embedded inside scattering and autofluorescing cellular environments, it is challenging to distinguish them against this background using conventional one-photon (i.e. linear) optical microscopy methods. Recently, we developed a multiphoton LM technique which exploits the four-wave mixing (FWM) nonlinearity of AuNPs, triply resonant to the LSPR. With this method we were able to detect individual small (down to 5\,nm radius) AuNPs inside scattering cells\,\cite{GiannakopoulouNS20,ZoriniantsPRX17} and tissues\,\cite{PopeSPIE21} completely free from background, at imaging speeds and excitation powers compatible with live cell imaging, with a sensitivity limited only by photon shot noise.   

Here, we demonstrate a CLEM workflow using individual small AuNPs as single probes of the epidermal growth factor (EGF) protein in mammalian cancer cells, imaged by FWM in LM and correlatively by transmission EM. Owing to the high photostability of AuNPs under ambient conditions, cryo-LM is not required in this workflow. To preserve the cellular ultrastructure and avoid artefacts from chemical fixation, we use vitrification by high-pressure freezing (HPF), followed by freeze substitution and resin embedding without additional heavy metal stains\,\cite{VerkadeJM08,vanWeeringMCB10}. Importantly, sections are imaged by FWM after sample preparation for EM, and a direct correlation with high accuracy is demonstrated using the very same AuNP observed under both modalities, without the need for additional fiducial markers.

\section{Results}\label{Results}

\subsection{Background-free four-wave mixing microscopy on EM-ready sections}\label{FWM}
\begin{figure*}[h]%
	\centering
	\includegraphics[width=0.9\textwidth]{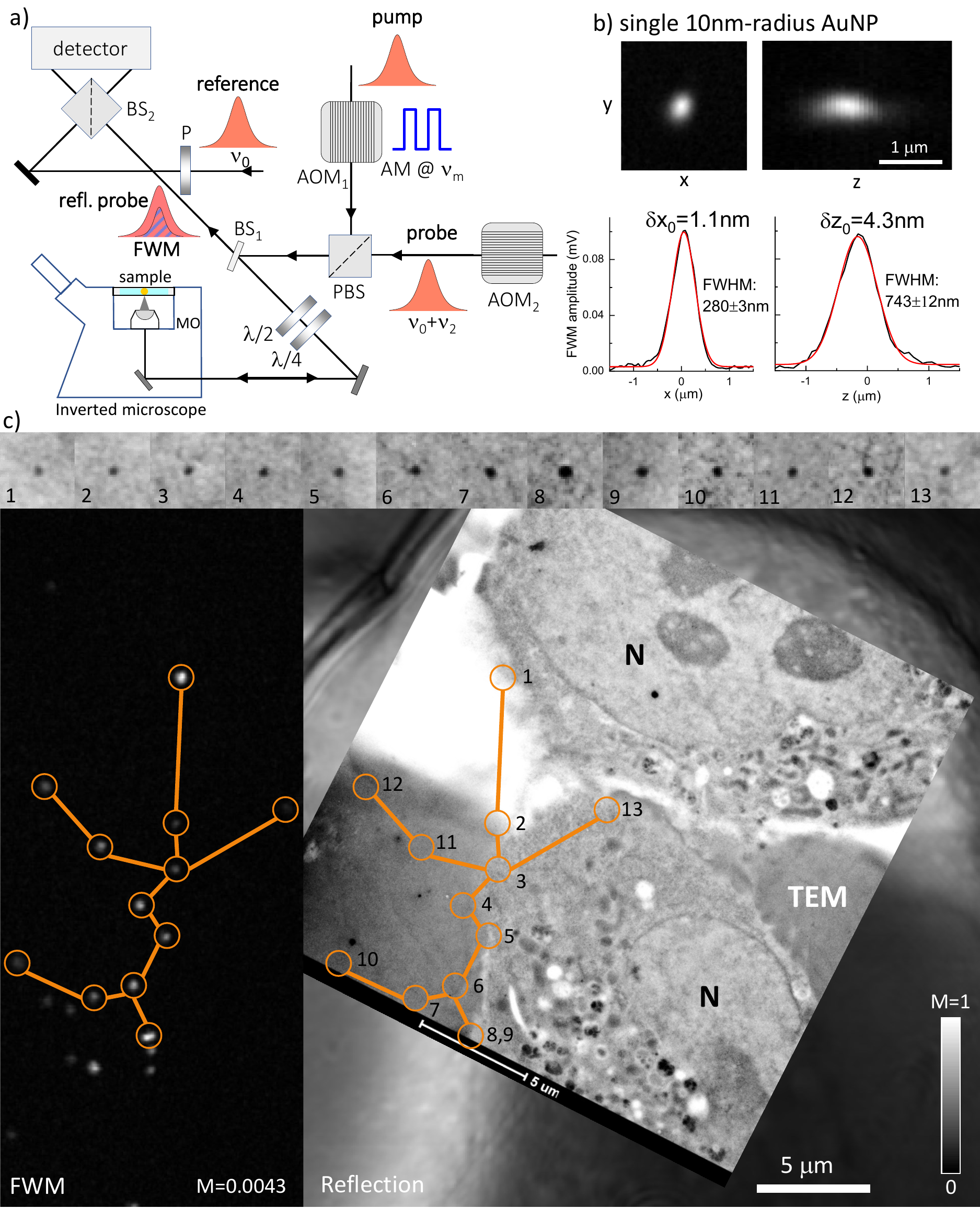}
	\caption{{\bf Correlative light-electron microscopy using FWM imaging}. a) Sketch of FWM set-up. Short optical pulses in resonance with the LSPR of AuNPs are focused onto the sample, using an inverted microscope, and generate a FWM field which is collected in epi-geometry, detected with a heterodyne interference scheme (see Methods). AOM: acousto-optic modulator. (P)BS: (polarising) beam splitter. P: polariser. MO: microscope objective. b) Example of volumetric FWM microscopy on a single 10\,nm-radius AuNP, with line-profiles along $x$ and $z$ at the $y$-position in the centre of the AuNP and corresponding Gaussian fits (red lines). The centroid localisation precision (${\delta}x_0$, ${\delta}z_0$) and the full-width at half maximum (FWHM) obtained from the fit are indicated. c) CLEM of 10\,nm-radius AuNPs bound to the EGF protein in HeLa cells. Individual AuNPs are detected background-free in FWM (left), measured directly on 300\,nm thick resin sections post-cell fixation, ready for EM analysis. The same pattern is found in TEM, highlighted by the orange circles. Two cells are visible, with their nucleus indicated (N). The nucleus is surrounded by the organelle-containing cytoplasm. The top row shows crops (0.2\,\textmu m$\times$0.2\,\textmu m) of the TEM image for each AuNP as numbered. The confocal reflection image simultaneously acquired with FWM is shown underneath the TEM image. Grey scales are from 0 to M as indicated (M=1 correspond to 31\,mV rms detected, see also Methods for details of the excitation and detection conditions).}\label{CLEM20nm}
\end{figure*}

In its general form, FWM is a third-order nonlinear light-matter interaction phenomenon wherein three light fields interact in a medium to generate a fourth wave. Here, we use a scheme where all waves have the same center frequency, and two of the incident light fields are identical (two-beam degenerate FWM). 
A sketch of the experimental setup implementing the FWM technique is shown in Fig.\,\ref{CLEM20nm}a. It exploits a combination of short optical pulses of about 150\,fs duration, called pump, probe and reference, generated by the same laser source (see also Methods). All pulses have the same center optical frequency, in resonance with the localised surface plasmon of nominally spherical small AuNPs. The detected FWM can be understood as a pump-induced change in the AuNP dielectric function, which manifests as a change in the scattering of the probe beam\,\cite{ZoriniantsPRX17}. Pump and probe pulses are focused onto the sample using a high numerical aperture (NA) microscope objective (MO), and the FWM signal is collected by the same objective in reflection (epi-geometry). To distinguish FWM from pump and probe beams, a heterodyne detection scheme is implemented, wherein the pump is amplitude modulated (at $\nu_{\rm m}$), the probe is radio-frequency shifted (by $\nu_{2}$), and the interference between FWM and reference fields is detected at the appropriate radio-frequency side-bands $\nu_{2} \pm \nu_{\rm m}$ (modulo the laser repetition rate\,\cite{ZoriniantsPRX17}). We have shown previously that the maximum FWM field amplitude is detected when the probe pulse arrives about 0.5\,ps after the pump pulse, which corresponds to the time needed for the free electron gas in the metal to reach the highest temperature (due the transfer of energy from the pump absorption) before starting to cool down via electron-phonon scattering\,\cite{MasiaPRB12}. As a result of this detection scheme, FWM is free from both linear scattering and incoherent (e.g. autofluorescence) background, and is temporally separated from instantaneous as well as long-lived non-linearities. Such exquisite background-free contrast is showcased in Fig.\,\ref{CLEM20nm}c, where FWM was acquired on AuNPs of nominal 10\,nm radius bound to the epidermal growth factor protein in HeLa cells, measured on 300\,nm thin sections ready for EM analysis, prepared using cell fixation by high-pressure freezing followed by freeze substitution and resin embedding (see Methods). Although these samples are embedded in LowicrylHM20 resin without addition of any electron dense staining agents, the sections create a strong background in the linear response, as shown in the confocal reflectance image acquired simultaneously with FWM in Fig.\,\ref{CLEM20nm}c. Yet, FWM is free from background and clearly shows the location of individual AuNPs (highlighted by the orange circles in Fig.\,\ref{CLEM20nm}c). The identical AuNP spatial pattern is found in the transmission EM (TEM) of the same section, correlatively measured after FWM imaging (see Methods), showcasing the suitability of AuNPs as single probes visible with high contrast in both FWM and EM. Notably, it is possible to locate the centroid position of single AuNPs in a FWM image with a localisation precision much better than the diffraction-limited spatial resolution, as shown in Fig.\,\ref{CLEM20nm}b. Gaussian fits of one-dimensional line-profiles along $x$ and $z$ at the $y$-position in the centre of a single AuNP provide a centroid localisation precision of about 1\,nm in plane and 4\,nm axially for the signal-to-noise ratio in the data. Furthermore, the FWM field phase in reflection encodes the axial displacement between particle and the focus center, thus it can be used to determine the particle $z$ coordinate without axial scanning\,\cite{ZoriniantsPRX17}. The linear dependence of the FWM phase versus $z$ measured on a set of AuNPs is reported on the Supplementary Information (SI) Fig.\,\ref{S-zdep}.    

\subsection{FWM is sensitive to the AuNP shape}\label{Shape}

It was shown in our previous work\,\cite{ZoriniantsPRX17} that using a polarisation-resolved configuration in the FWM field detection provides additional information on the AuNP shape and orientation. In this configuration, probe and pump beams, linearly polarised in the
laboratory system, are transformed into circularly polarised
beams at the sample by a combination of $\lambda/4$ and $\lambda/2$ waveplates (see also Fig.\,\ref{CLEM20nm}a). We then use a dual-polarisation balanced detection (see Methods) which allows us to detect the co- and cross-circularly polarised components of the reflected probe and FWM fields relative to the
incident circularly polarized probe, having amplitudes (phases)
indicated as $\Ar^\pm$ and $\AF^\pm$ ($\Phir^\pm$ and $\PhiF^\pm$), respectively,
where $+$ ($-$) refers to the co (cross) polarised component. Notably, we found, with the aid of numerical simulations of the detected FWM field spatial pattern compared with the experiments, that the cross-polarised component is strongly sensitive to small AuNP shape asymmetries, which are always present in these nominally-spherical AuNPs consistent with their morphology observed in TEM. Using an ellipsoid model to account for deviations from spherical shapes, the calculations showed that the amplitude ratio $\AF^- / \AF^+$ at the AuNP center is proportional to the AuNP ellipticity, and that the phase $\PhiF^- - \PhiF^+$ reports the in-plane particle orientation\,\cite{ZoriniantsPRX17}. 

\begin{figure*}[h]%
 	\centering
 	\includegraphics[width=1\textwidth]{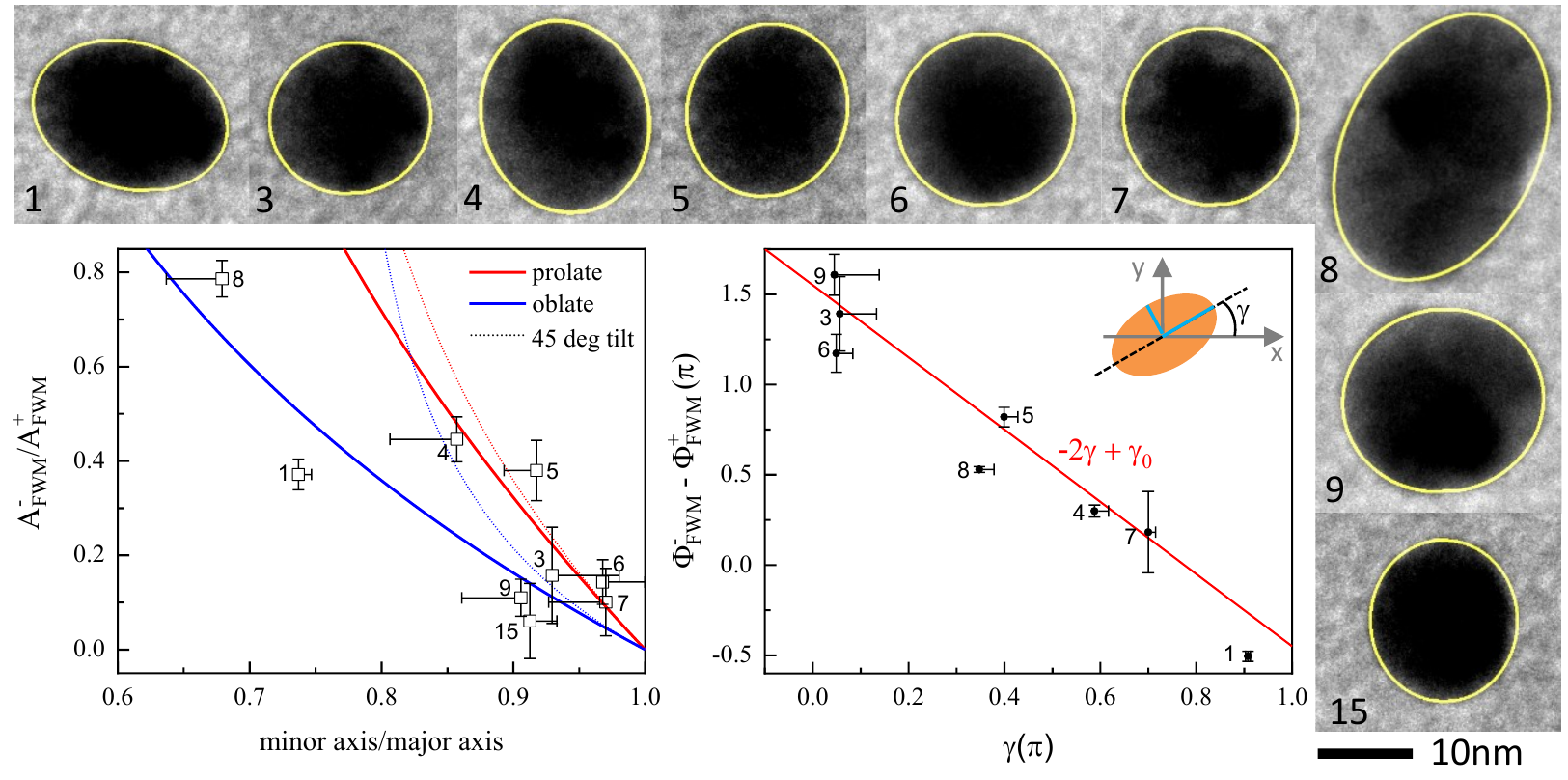}
 	\caption{{\bf FWM dependence on the AuNP aspect ratio and orientation}. High-magnification TEM images of selected nanoparticles (numbered as in Fig.\,\ref{CLEM20nm}c) were fitted with an elliptical shape (shown in yellow). AuNP 15 is not captured by the TEM overview in Fig.\,\ref{CLEM20nm}c and is shown in SI Fig.\,\ref{S-CLEMadd} and Fig.\,\ref{S-20nm_correlation_ED}. The corresponding major and minor axis and the orientation angle $\gamma$ were obtained (see also sketch). The ratio of the cross- to co-circularly polarised FWM components in amplitude ($\AF^- / \AF^+$) and phase ($\PhiF^- - \PhiF^+$) is shown versus the in-plane elliptical aspect ratio and orientation angle. Vertical error bars represent the single-pixel shot noise in the FWM measurements. Single-sided horizontal bars were obtained by fitting the TEM images with a variable contrast threshold (see text). For $\AF^- / \AF^+$, red (blue) lines are calculated dependences assuming a prolate (oblate) ellipsoid with axis $a>b=c$ ($a<b=c$) aligned along the $x, y, z$ directions respectively. Dotted lines assume that the NP $a$ and $c$ axis are tilted by 45 degrees in the $x,z$ plane, having a projected axis in the $x$-direction as derived in Section\,\ref{S-TEMellipse_model}. }\label{ellipticity}
 \end{figure*}

Using the CLEM workflow, here we have correlatively analysed the measured FWM field ratio and the AuNP shape obtained with TEM, and compared the results with the ellipsoid model previously developed. Fig.\,\ref{ellipticity} shows high-magnification TEM images on a selection of the AuNPs seen in Fig.\,\ref{CLEM20nm}c, as indicated by the corresponding numbers. An ellipse was fitted to these images as shown by the yellow lines (see also Methods). The corresponding major and minor axis and the orientation angle $\gamma$ were obtained (see sketch in Fig.\,\ref{ellipticity}) and the dependence of the measured FWM field ratio at the AuNP center is shown in the plots, for both amplitude and phase components. Error bars in the measured FWM field ratio represent the shot-noise in regions away from the AuNPs while the horizontal error bars were obtained by changing the threshold levels used to fit an ellipse to the TEM images (see Methods). For this analysis, we ensured that the selected NPs were sufficiently in focus (see SI Fig.\,\ref{S-ratiovsz}), to justify comparing the experimentally measured FWM ratio with the ellipsoid model. The latter was developed assuming a prolate or an oblate NP shape, with semi-axis $a>b=c$ or $a<b=c$ along the $x, y, z$ directions, respectively. We also considered the case of a tilted ellipsoid rotated by 45 degrees in the $x,z$ plane, and calculated the projected semi-axis along $x$ accordingly (see SI section\,\ref{S-TEMellipse}). The corresponding amplitude ratios $\AF^- / \AF^+$ derived from such model are shown in Fig.\,\ref{ellipticity} as labelled. Generally, the experimental data agree well with the model, taking into account that the TEM used here is an in-plane projection of the 3D shape, hence we cannot tell if a NP is oblate or prolate and how its axes are orientated. Notably, NPs number 1 and 15 show a darker contrast in TEM, consistent with having an oblate shape with the long c-axis out of plane. Regarding the NP in-plane orientation, the experimental FWM ratio phase $\PhiF^- - \PhiF^+$ exhibits a good agreement with the dependence $-2\gamma + \gamma_0$, where $\gamma_0$ is a rotation offset, as predicted by the ellipsoid model\,\cite{ZoriniantsPRX17}.

\subsection{FWM-EM correlation accuracy}\label{CLEMaccuracy}

The correlation accuracy between locating the same AuNP in FWM and in TEM was evaluated as follows. The centroid ${\bf r_0}=(x_0,y_0)$ position coordinates of each individual AuNP in a FWM image were obtained using a two-dimensional Gaussian fit of the $\AF^+$ profile (see Methods). The corresponding AuNP coordinates in the EM were assigned by examining zooms at the particle location and positioning the particle centre based on shape geometry. The two sets of coordinates were then compared using a linear transformation matrix. Specifically, the coordinates of each AuNP in the FWM image were transformed into the reference system of the EM using an affine transformation $\mathbf{C}$, including shear, scaling, rotation and translation, so that ${\bf r_B}=\mathbf{C}({\bf r_A})$ where ${\bf r_A}$ is the coordinate vector in the FWM image and ${\bf r_B}$ is the coordinate vector in the EM image. For more than 3 AuNPs, the system is overdetermined and $\mathbf{C}$ is calculated by minimising the sum of the squared deviations over all particle coordinates (see also Methods). As a measure of the correlation accuracy we then evaluate the quantity $S=\sqrt{\frac{1}{N}\sum_i \|{\bf r_B}_i-\bf{C}({\bf r_A}_i)\|^2}$ where $N$ is the total number of AuNPs being compared and $i=1,....,N$ denotes the individual i-th particle.
  
\begin{figure*}[h]%
	\centering
	\includegraphics[width=1\textwidth]{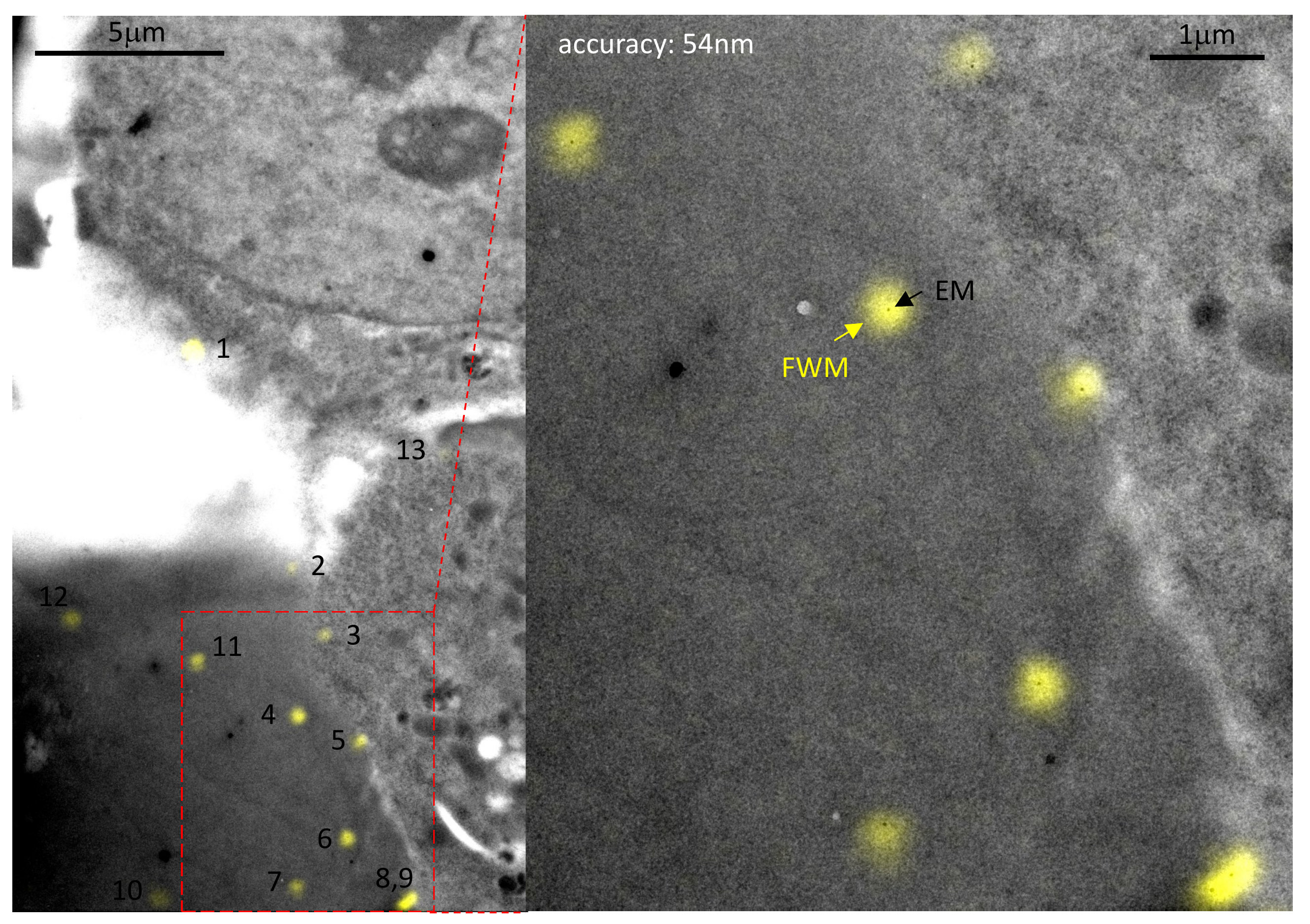}
	\caption{{\bf CLEM correlation accuracy.} Overlay of FWM field amplitude (yellow) and TEM image (grey) from Fig.\,\ref{CLEM20nm}c (contrast adjusted for visibility). The FWM image is transformed into the EM reference system using a linear transformation matrix that accounts for translation, rotation, shear and scaling (see text). The correlation accuracy is indicated.}\label{20nm_correlation}
\end{figure*}

\begin{figure*}[h]%
	\centering
	\includegraphics[width=\textwidth]{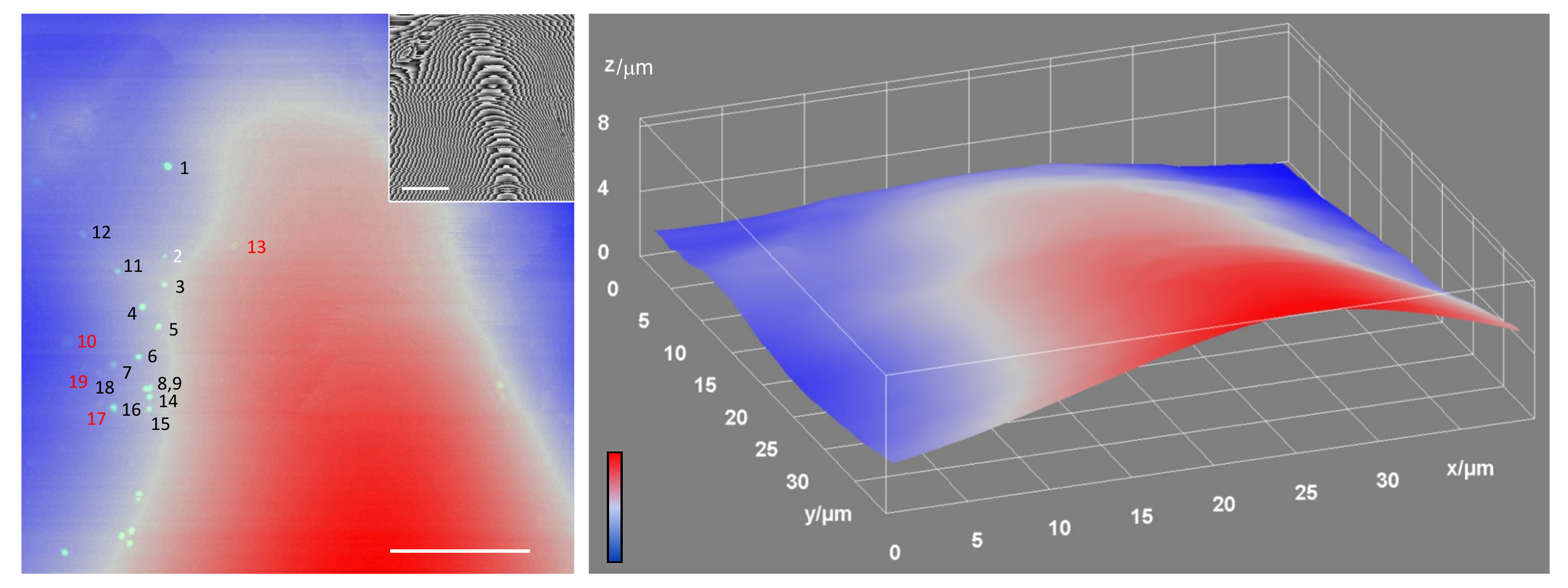}
	\caption{{\bf Surface height profile of resin section in FWM imaging} Colour-coded height profile (blue: 0, red: 8.51\,\textmu m) obtained from the phase of the reflected probe field, shown in the inset on a grey scale from $-\pi$ (black) to $\pi$ (white), for the region in Fig.\,\ref{CLEM20nm}c and Fig.\,\ref{S-20nm_correlation_ED}. The FWM field amplitude of AuNPs is overlaid on the left panel and AuNPs are labelled. The surface profile shows height differences of several microns, indicating a ripple in the pioloform layer and the supported 300\,nm thin resin section. AuNPs labelled in red are those excluded from the CLEM correlation analysis as they are too out of focus (AuNP 10 is 1.9\,\textmu m below and AuNP 13 is 0.8\,\textmu m above AuNP 5). Scale bar: 10\,\textmu m.}\label{surface}
\end{figure*}

Fig.\,\ref{20nm_correlation} shows an example of this analysis for the 10\,nm-radius AuNPs reported in Fig.\,\ref{CLEM20nm}c, where the transformed FWM image has been overlaid to the EM image. A zoom of the overlay is given in Fig.\,\ref{20nm_correlation} to showcase the overlap between an individual AuNP in FWM (yellow spots) and EM (black spots). A correlation accuracy of 94\,nm is found when including all numbered particles in Fig.\,\ref{CLEM20nm}c, which reduces to 54\,nm when excluding AuNPs 2, 10 and 13. These three particles have a FWM amplitude below a third of the typical maximum value observed. A high-magnification TEM inspection of AuNP 2 shows a weak contrast (see SI  Fig.\,\ref{S-TEMAuNP2}), hence an atypical structure, while AuNP 10 and 13 have a low FWM amplitude because they are significantly out of focus, as demonstrated by an analysis of the point-spread-function (PSF) width and the error in the centroid localisation precision (see SI Section\,\ref{S-CLEMaccuracy}, Fig.\,\ref{S-PSF20nm}). 
When particles are out of focus, not only the localisation precision decreases but their location is also affected by additional uncertainties, including objective aberrations and deformations of the pioloform layer supporting the resin section which change from FWM in water to EM in vacuum (see Methods). Notably, by exploiting the topography information encoded in the detected phase of the reflected probe field, we reconstructed a height profile of the resin section for the region in Fig.\,\ref{CLEM20nm}c, showing that there is a vertical tilt/bending of the pioloform layer, and AuNP 10 and 13 are indeed located at significantly different heights compared to the other particles (see SI Section\,\ref{S-CLEMaccuracy}, and Fig.\,\ref{surface} which shows AuNP 10 being 1.9\,{\textmu}m below and AuNP 13 being 0.8\,{\textmu}m above AuNP 5). This also explains why, despite the resin section being only 300\,nm thick, hence smaller the axial extension of the PSF in FWM imaging (as shown in Fig.\,\ref{CLEM20nm}b), we do have issues of AuNPs being out of focus.  

It should be highlighted that a correlation accuracy of 54\,nm is remarkably small considering the large size ($>10$\,{\textmu}m) of the region over which the correlation is carried out. An additional example using a different, slightly smaller, EM region (centred around AuNPs 8 and 9) is shown in Fig.\,\ref{S-20nm_correlation_ED}, giving a correlation accuracy of 43\,nm, when excluding AuNP 17 and 19 from the analysis after consistently applying the same out-of-focus criteria mentioned above (see SI Section\,\ref{S-CLEMaccuracy}, Fig.\,\ref{S-PSF20nm} for details).        

\begin{figure*}[h]%
	\centering
	\includegraphics[width=1\textwidth]{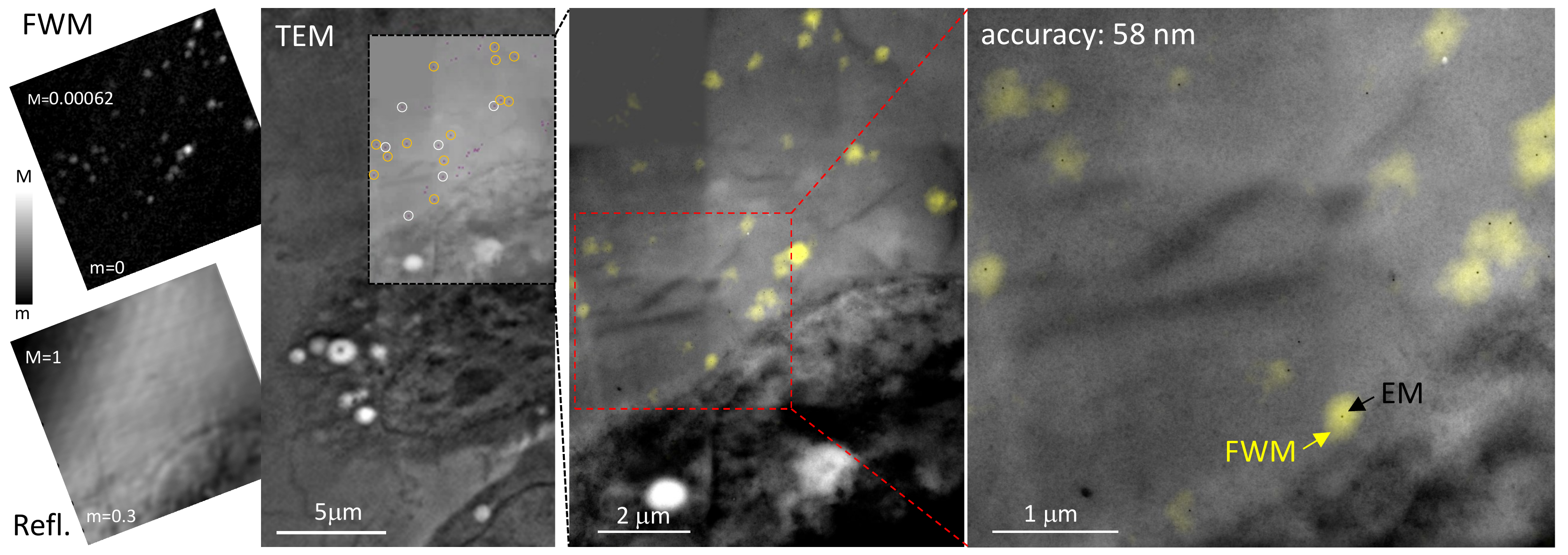}
	\caption{{\bf CLEM correlation accuracy with 5nm-radius AuNPs}. HeLa cells incubated with 5nm-radius AuNPs bound to the EGF protein. Individual AuNPs are detected background-free in FWM (left) measured on 300\,nm thin resin sections post-cell fixation, ready for EM analysis. The confocal reflection image simultaneously acquired with FWM is shown below (linear grey scales are from m to M as indicated; M=1 correspond to 65\,mV rms detected, see Methods for details of the excitation and detection conditions). A large overview TEM of the same region is shown. On the area indicated by the black dashed frame, a series of high resolution EM images were taken and stitched together. Individual AuNPs are highlighted by the circles. The overlay between FWM (yellow) and TEM (grey) is shown on the center and further zoomed into the indicated red dashed area on the right side (contrast adjusted to aid visualisation). For the correlation analysis, of the 19 individual AuNPs highlighted by the circles, 6 (white circles) were discarded as being of focus. The FWM image was transformed into the EM reference system using a linear transformation matrix that accounts for translation, rotation, shear and scaling of axes. On the right side, individual AuNPs identified in FWM (yellow spots) are seen in EM (black dots). The correlation accuracy is indicated. }\label{10nm_correlation_ED}
\end{figure*}

We also investigated HeLa cells incubated with 5\,nm-radius AuNPs. It was shown in our previous work\,\cite{MasiaPRB12} that the FWM field amplitude scales almost proportionally with the AuNP volume. Therefore, the signal to noise ratio, and in turn the localisation precision, is decreased by about 8-fold compared to using 10nm-radius AuNPs under identical excitation and detection conditions. Still, individual nanoparticles of this small size can be clearly resolved in FWM microscopy, above noise and background-free, as we showed in Ref.\,\cite{GiannakopoulouNS20}. An example of CLEM with FWM imaging using 5\,nm-radius AuNPs in HeLa cells is shown in Fig.\,\ref{10nm_correlation_ED}. Several AuNPs are clearly visible in both FWM and TEM. A few AuNPs are too close to be spatially distinguished in the FWM image, but 19 individual AuNPs are available for position analysis. This resulted in a correlation accuracy of 58\,nm, whereby 13 individual AuNPs were used for the correlation (see orange circles in Fig.\,\ref{10nm_correlation_ED}), and 6 nanoparticles were excluded (white circles in Fig.\,\ref{10nm_correlation_ED}) based on the  out-of-focus criteria discussed previously (see SI Section.\ref{S-CLEMaccuracy}, Fig.\,\ref{S-PSF10nm}).  Another example showing an adjacent region is provided in  Fig.\,\ref{S-CLEM10nm}. Merging both regions results in a correlation accuracy of 63\,nm (see SI section\,\ref{S-CLEMaccuracy}). 

We should note that the value $S$ scales with the number of particles included in the analysis $N$ and the number $M$ of parameters in the transformation according to $\sqrt{(2N-M)/(2N)}$. In other words, decreasing the number of particles in the analysis decreases the quantity $S$ (as stated above, if $N$=3 the M=6 parameters of $\mathbf{C}$ are fully determined from linear algebra and $S=0$). To account for this, we can calculate a corrected correlation accuracy as $S/\sqrt{1-(M/2N)}$. This is found to be 65\,nm both for the 10\,nm-radius AuNPs in Fig.\,\ref{20nm_correlation} and for the 5\,nm-radius AuNPs in Fig.\,\ref{10nm_correlation_ED}.      
 
\begin{figure*}[h]%
	\centering
	\includegraphics[width=1\textwidth]{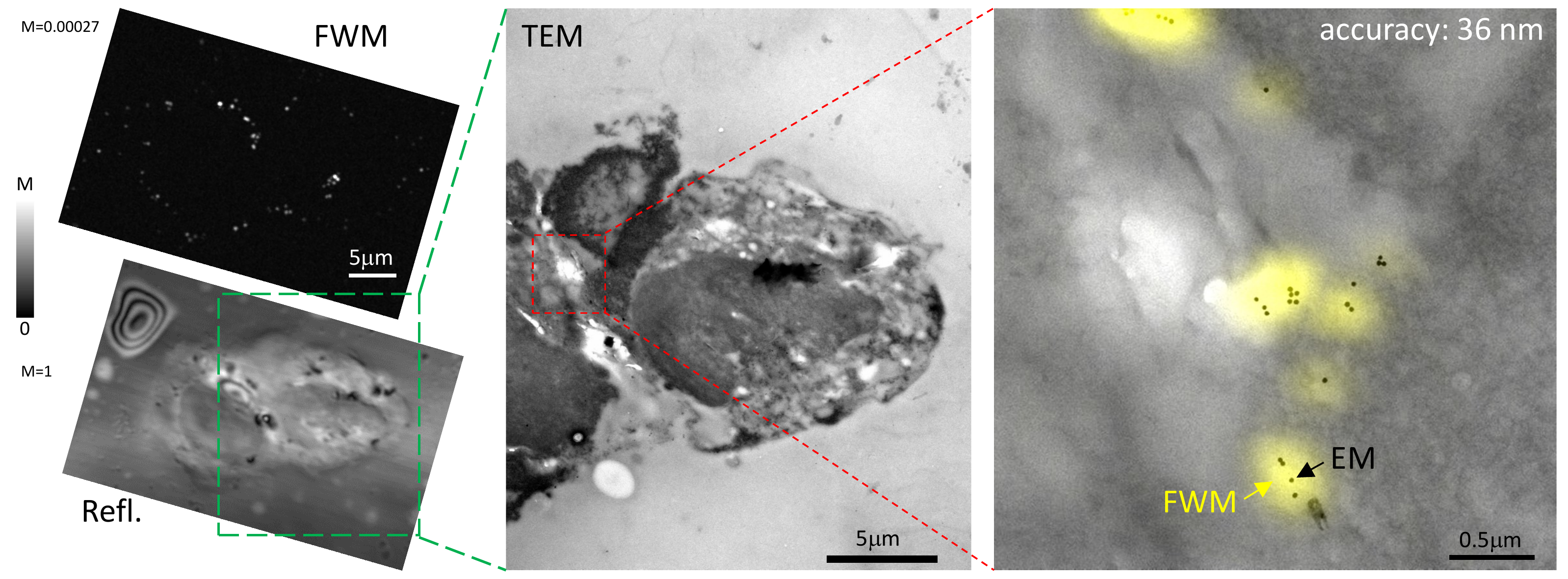}
	\caption{{\bf CLEM correlation accuracy with 3D FWM analysis.} FWM-CLEM using 10nm-radius AuNPs bound to EGF internalised in HeLa cells whereby the coordinates of the particles in FWM are measured in 3D via a z-stack. A large FWM overview in 2D with corresponding reflection image measured simultaneously is shown on the left (linear grey scales are from 0 to M as indicated; M=1 corresponds to 33mV rms detected; see Methods for details of the excitation and detection conditions). A TEM overview of the same region is shown in the center, as indicated by the green dashed frame. On the left, an overlay of FWM field amplitude (yellow) and TEM image (grey) is shown for the region highlighted by the red dashed frame, where FWM is a maximum amplitude projection from a 3D z-stack (50\,nm step size in $z$). AuNPs form small clusters and are no longer resolved as individual particles in FWM. The centroid position of each cluster was determined in 3D from the FWM z-stack (see Methods), and its 2D in-plane coordinates were compared with the position of the geometrical centre of the cluster in TEM (which is a 2D transmission projection) for the correlation analysis. The resulting correlation accuracy from the comparison of the six clusters shown in the figure is indicated.}\label{20nm_cluster}
\end{figure*}

Considering that the shot-noise limited precision in locating the centroid position of a AuNP in focus by FWM is only a few nanometres (see Fig.\,\ref{CLEM20nm}b), the measured values of $S$, even after excluding AuNPs which are too out of focus, are limited by systematic errors, i.e. $S$ is dominated by accuracy rather than precision. To address this point, we performed FWM-CLEM using 10\,nm-radius AuNPs  whereby the coordinates of the particles in FWM were measured in 3D with a fine axial scan (50\,nm step size in $z$), such that the coordinates at the plane of optimum focus are accurately determined and systematics from e.g. out-of-focus aberrations are eliminated. These results are summarized in Fig.\,\ref{20nm_cluster}. Notably, here we observe AuNPs which have been internalised inside the cells (instead of being outside or at the cell surface, as in Fig.\,\ref{CLEM20nm}c). AuNPs form small clusters and are no longer resolved as individual particles in FWM. Therefore, in this case, we determined the centroid position of the cluster in 3D from the FWM z-stack (see Methods), and compared its 2D in-plane coordinates with the position of the geometrical centre of the cluster in TEM (which is a 2D transmission projection) for the correlation analysis. The resulting correlation accuracy for the six clusters shown in Fig.\,\ref{20nm_cluster} is 36\,nm. Another example correlating 10 clusters is provided in Fig.\,\ref{S-SM20nm_cluster}, for which an accuracy of 44\,nm is found.

\section{Discussion}\label{Discussion}

The demonstration of FWM-CLEM with a single AuNP probe opens new possibilities for correlative light electron microscopy workflows. As shown here, we can locate the position of a single AuNP with nanometric precision at ambient conditions, without the need for cryo light microscopy, owing to the background-free and photostable FWM response of individual AuNPs which do not photobleach. The very same AuNP is well visible in EM due to its electron dense composition, offering high correlation accuracy without the need for additional fiducials. We have shown proof-of-principle results with 10\,nm-radius and 5\,nm-radius AuNPs bound to the EGF protein in HeLa cells, using FWM directly on 300\,nm thin sections prepared for EM by high pressure freezing, freeze substitution and LowicrylHM20 resin embedding without using heavy metal stains. 

Generally, we found a correlation accuracy limited by systematics, in the range of 60\,nm or less over areas larger than 10\,{\textmu}m. Systematic errors included a bending of pioloform layer supporting the resin section, which changes from FWM in water to TEM in vacuum. This is difficult to correct for by  coordinate transformations, and likely to require non-trivial methods beyond the linear transformation used by us. Importantly, systematics can be improved in future experimental designs, such that a correlation uncertainty eventually limited only by localisation precision from photon shot-noise, and hence down to 5-10\,nm (or even lower by measuring longer, considering the photostability of AuNPs) could be reached. Moreover, since a single probe is used, as soon as this is identified from FWM into the TEM image, its relationship with the cellular ultrastructure is unambiguously determined. We should also highlight that FWM is compatible with live cell imaging\,\cite{PopeSPIE21}, hence can be applied from the start of a CLEM workflow, before cell fixation, as well as post fixation. 

The detection of individual AuNPs with FWM lends itself to applications in single particle tracking (SPT) inside living cells\,\cite{ManzoRPP15}, to follow e.g. the entry and intracellular pathways of single molecules tagged with AuNPs, from proteins to drugs. A related application is following the fate of individual virions\,\cite{LiuCR20} to gain spatio-temporal insights into fundamental mechanisms of virus transport and infection occurring in live cells. Combined with existing strategies to label with or even encapsulate AuNPs inside virions\,\cite{ZhangNT17}, FWM opens the exciting prospect to track single virions over long observation times, background-free and deep inside living cells and tissues, to then pin-point events of interest (e.g. genome release) in the context of the cellular ultrastructure by CLEM.    

While in the present demonstration we have shown AuNPs probes down to 5\,nm radius, we emphasise that smaller probes could be used. In fact, in our previous work\,\cite{MasiaOL09} we reported FWM microscopy with 2.5nm radius AuNPs immunostaining the Golgi apparatus of HepG2 cells, where nanoparticles were detected as clusters in the focal volume. The FWM field amplitude scales proportionally with the NP volume and with the number of isolated particles in the focal volume, thus 8 AuNPs of 2.5nm-radius provide the same FWM signal as a single 5nm-radius AuNPs under the same excitation and detection conditions. The FWM amplitude signal-to-noise ratio scales as $\sqrt{t}I_1\sqrt{I_2}$ with $I_1$ ($I_2$) being the intensity of the pump (probe) beam at the sample and $t$ the integration time \,\cite{ZoriniantsPRX17}, hence to detect a single 2.5\,nm radius AuNP (instead of a cluster) one can increase the excitation power and integration time accordingly.  However, these conditions might prevent the applicability of the technique to living cells, due to nanoparticle heating under high power illumination and/or integration times becoming too long for the dynamics under observation. Alternatively, 2.5\,nm radius silver nanoparticles can be used, as these have a 10-fold larger polarisability compared to a AuNP of equal radius (and correspondingly will exhibit higher FWM), owing to their sharper LSPR in the absence of interband transitions, as was seen in their photothermal response\,\cite{BerciaudPRB06}.   

Another interesting consideration is the sensitivity of polarisation-resolved FWM to the shape and orientation of individual AuNPs, as we have shown here and previously\,\cite{ZoriniantsPRX17}. From a single particle tracking standpoint, this opens the exciting prospect of tracking particle rotations as well as translations, while for imaging it provides an opportunity for multiplexing by size and shape recognition. Finally, we highlight the recent demonstration that AuNPs can be synthesised directly inside cells and attached to specific biomolecules using genetic tagging\,\cite{JangNM20}. This could bring a "bioimaging revolution" to FWM microscopy and FWM-CLEM, similar to the advent of fluorescent proteins in fluorescence microscopy.



\section{Methods}\label{Methods}
{\bf FWM set-up.} FWM microscopy was performed using a home built set-up, as described in detail in our recent works\,\cite{GiannakopoulouNS20,ZoriniantsPRX17}.  Briefly, optical pulses of 150\,fs duration centered at 550\,nm wavelength
with $\nu_{\rm L}$=80\,MHz repetition rate were provided by the
signal output of an optical parametric oscillator (Spectra Physics
Inspire HF 100) pumped by a frequency-doubled femtosecond Ti:Sa
laser (Spectra Physics Mai Tai HP). The output was split into three
beams having the same center optical frequency, resulting in a
triply degenerate FWM scheme. One beam acts as a pump and excites
the AuNP at the LSPR, with an intensity that is modulated
at $\nu_{\rm m}$=0.4\,MHz by an acousto-optic modulator (AOM). The
change in the AuNP optical properties induced by this excitation is
resonantly probed by a second pulse at an adjustable delay time after the pump pulse. Pump and probe pulses are recombined
into the same spatial mode and focused onto the sample by a
60$\times$ water-immersion objective of 1.27\,NA (Nikon CFI Plan
Apochromat lambda series MRD70650) mounted onto a commercial inverted microscope (Nikon Ti-U) with a 1.5$\times$ tube lens. The sample is positioned and moved with
respect to the focal volume of the objective by scanning a $xyz$
sample stage with nanometric position precision (MadCityLabs
NanoLP200). A FWM field (proportional to the pump induced change of
the probe reflected field) is collected by the same objective
(epi-detection), together with the probe reflected field,
transmitted by an 80:20 (T:R) beam splitter (BS$_1$ in Fig.\,\ref{CLEM20nm}) used to couple the incident
beams into the microscope, and recombined in a 50:50 beam splitter (BS$_2$)
with a reference pulse field of adjustable delay. The resulting
interference is detected by two pairs of balanced Si photodiodes
(Hamamatsu S5973-02). A heterodyne scheme discriminates the FWM
field from pump and probe pulses and detects the amplitude and phase
of the field. In this scheme, the probe optical frequency is upshifted by a radio frequency amount ($\nu_{2}$=82\,MHz),
and the interference of the FWM with the unshifted reference field
is detected. As a result of the amplitude modulation of the pump at
$\nu_{\rm m}$ and the frequency shift of the probe by $\nu_2$, this
interference gives rise to a beat note at $\nu_2$, with two
sidebands at $\nu_2\pm\nu_{\rm m}$, and replica separated by the
repetition rate $\nu_{\rm L}$ of the pulse train frequency comb. A multi-channel
lock-in amplifier (Zurich Instruments HF2LI) enables the
simultaneous detection of the carrier at $\nu_2-\nu_{\rm L}$=2\,MHz
and the sidebands at $\nu_2\pm\nu_{\rm m}-\nu_{\rm
	L}=2{\pm}0.4$\,MHz. As described in our previous
work\,\cite{ZoriniantsPRX17} the set-up also features a dual
polarization scheme. Briefly, in this scheme, probe and pump beams, linearly
polarised horizontally (H) and vertically (V) respectively in the
laboratory system, are transformed into cross-circularly polarized
beams at the sample by a combination of $\lambda/4$ and $\lambda/2$
waveplates (see Fig.\,\ref{CLEM20nm}a). The reflected probe and FWM fields collected by the 
microscope objective travel backwards through the same waveplates,
such that the probe reflected by a planar surface returns V polarized in the laboratory system.
The reference beam is polarised at 45 degree (using a polariser) prior to recombining with the
epi-detected signal via the non-polarizing beamsplitter BS$_2$. A
Wollaston prism vertically separates H and V polarizations for each
arm of the interferometer after BS$_2$. Two pairs of balanced
photodiodes then provide polarization resolved detection, the bottom
(top) pair detecting the current difference (for common-mode noise
rejection) of the V (H) polarised interferometer arms. The measured interference corresponds to the co- and cross-circularly polarised
components of the reflected probe and FWM fields relative to the
incident circularly polarized probe, having amplitudes (phases)
indicated as $\Ar^\pm$ and $\AF^\pm$ ($\Phir^\pm$ and $\PhiF^\pm$), respectively,
where $+$ ($-$) refers to the co (cross) polarised component.

The results in Fig.\,\ref{CLEM20nm}b,c refer to the co-polarised component and the acquisition parameters were as follows: pump-probe delay time of 0.5 ps, b) pump (probe) power at the sample
of 100\,\textmu W (50\,{\textmu}W), 3\,ms-pixel dwell time, pixel size
in plane of 21\,nm and z stacks over 3\,\textmu m in 75\,nm z steps; c) pump (probe) power at the sample
of 80\,\textmu W (40\,\textmu W), 1\,ms-pixel dwell time, pixel size
in plane of 72\,nm. The FWM is shown as a maximum amplitude projection for two $xy$ planes 0.5\,\textmu m separated in $z$. 

The results in Fig.\,\ref{10nm_correlation_ED} refer to the co-polarised component and the acquisition parameters were as follows: pump-probe delay time of 0.5 ps, pump (probe) power at the sample
of 100\,\textmu W (50\,\textmu W), 3\,ms-pixel dwell time, pixel size
in plane: 43\,nm. 

The results in Fig.\,\ref{20nm_cluster} refer to the co-polarised component and the acquisition parameters were as follows. 2D overview: pump-probe delay time of 0.5 ps, pump (probe) power at the sample
of 20\,\textmu W (10\,\textmu W), 1\,ms pixel dwell time, pixel size
in plane 72\,nm. 3D stack: pump-probe delay time of 0.5 ps, pump (probe) power at the sample
of 20\,\textmu W (10\,\textmu W), 1\,ms pixel dwell time, pixel size
in plane 80\,nm, 50\,nm step size in $z$ and 61 z-steps (3\,\textmu m total range). 

{\bf Sample preparation.} HeLa cells were grown on 1.5 mm wide sapphire discs (Leica Microsystems)\,\cite{VerkadeJM08}. Following a 2-hour serum starvation, EGF-coupled to 5 or 10\,nm radius AuNP was allowed to internalise into the HeLa cells for 20\,minutes\,\cite{BrownProtoplasma2010}.  After a brief rinse in 20\% BSA in growth medium, the disc was placed in a 0.1\,mm deep membrane carrier and high pressure frozen (EMPACT2 + RTS, Leica Microsystems)\,\cite{VerkadeJM08}. The frozen carrier was transferred under liquid nitrogen to an automated freeze substitution device (AFS2 + FSP, Leica Microsystems). Freeze substitution to Lowicryl HM20 was performed as described in \cite{vanWeeringMCB10} with the exception that any heavy metal stain was omitted. Following UV polymerisation of the resin, 
300\,nm resin sections were cut and mounted onto copper slot grids on a layer of pioloform. For FWM imaging, the copper grids were mounted in water between a glass coverslip (Menzel Gläser, 24\,mm$\times$24\,mm, \# 1.5) and a slide (Menzel Gläser, 76\,mm$\times$26\,mm$\times$1.0\,mm) inside a 0.12\,mm thick (13\,mm chamber diameter) imaging gasket (Grace Bio-Labs, SecureSeal$^{\rm TM}$). The copper grid was orientated such that the 300\,nm sections were facing the coverslip.

{\bf Data analysis.}
The experimental shot noise was
evaluated by taking the statistical distribution of the
measured FWM field (both in the in-phase and in-quadrature components detected by the lock-in amplifier) in a spatial region where no FWM is present. The standard
deviation of this distribution was deduced and was found
to be identical in both components, as well as for the co-polarised
and cross-circularly polarised components, as expected for
an experimental noise dominated by the shot noise in the
reference beam\,\cite{ZoriniantsPRX17}. The error bars in the FWM field ratio in Fig.\,\ref{ellipticity} are calculated by propagating the errors from the experimental shot noise in the co- and cross-circularly polarised components, and are shown as two standard deviations. The FWM field ratios in Fig.\,\ref{ellipticity} were measured from the two in-plane data sets 0.5\,\textmu m apart in $z$ forming the overview in Fig.\,\ref{CLEM20nm}c. Notably, the FWM ratio values are slightly dependent on the axial position of the AuNP. Hence, care was taken to consider the ratio only for NPs that were in focus, based on the maximum co-polarised FWM amplitude detected and on the width of the point-spread function (see SI Fig.\,\ref{S-ratiovsz}).  

The fitted ellipses to the TEM images in Fig.\,\ref{ellipticity} are obtained using the "Analyse particles - fit ellipse" command in the freely available Java-based image analysis program ImageJ\,\cite{ImageJ}. This command measures and fits objects in thresholded images. It works by scanning the selection until it finds the edge of an object. It then provides the major and minor semi-axis and the orientation angle $\gamma$ of the best fitting ellipse. The orientation angle is calculated between the major axis and a line parallel to the $x$-axis of the image (see sketch in Fig.\,\ref{ellipticity}). For the ellipses shown by the yellow lines in the TEM images in Fig.\,\ref{ellipticity}, the "auto-threshold" default option was applied. To estimate the error bars in the fitted aspect ratios and in the angle $\gamma$, TEM images were re-fitted using a different threshold such that the area of the fitted ellipse was 80\% of the area obtained with auto-threshold, as shown in SI Fig.\,\ref{S-ellipse}. The horizontal errors bars in Fig.\,\ref{ellipticity} are the single-sided distances between the values using the auto-threshold option and the re-fitted values.  

{\bf Centroid fitting.} To determine the centroid position of the NPs, we have fitted the spatially resolved FWM field with a Gaussian complex function given by
\begin{equation}
\begin{split}
\mathfrak{G}\left(x,y\right)=G e^{\iota \phi} {\rm exp}\Bigg[\Bigg.-\frac{4\log{(2)}}{w^2}\\
\Bigg(\Bigg.\epsilon\bigg(\left(x-x_0\right)\cos{\theta}-\left(y-y_0\right)\sin{\theta}\bigg)+ \\
(1/\epsilon)\bigg(\left(x-x_0\right)\sin{\theta}-\left(y-y_0\right)\cos{\theta}\bigg)\Bigg.\Bigg)^2\Bigg.\Bigg],
\end{split}
\end{equation}
where $G$ is the amplitude of the signal at the peak, $\phi$ its phase, $w$ a mean width of the peak, $x_0$ and $y_0$ the coordinates of the centroid, $\epsilon$ the ellipticity of the peak and $\theta$ the orientation.

{\bf Affine transformation.}
We use the linear transformation between the coordinates of image A and the coordinates of image B 
\begin{equation}
{\bf r_B}=\mathbf{C} ({\bf r_A})=\mathbf{H}\mathbf{S}\mathbf{R}{\bf r_A}+\mathbf{T},
\end{equation}
with the shear ($\mathbf{H}$), scaling ($\mathbf{S}$), rotation ($\mathbf{R}$) and translation ($\mathbf{T}$),
given by
\begin{equation}
\begin{split}
\mathbf{H}&=
\begin{pmatrix}
1 & h \\
0 & 1  
\end{pmatrix} \\
\mathbf{S}&=
\begin{pmatrix}
s_{\rm x} & 0 \\
0 & s_{\rm y} \\
\end{pmatrix}\\
\mathbf{R}&=
\begin{pmatrix}
\cos{\alpha} & -\sin{\alpha} \\
\sin{\alpha} & \cos{\alpha} \\
\end{pmatrix}\\
\mathbf{T}&=
\begin{pmatrix}
t_{\rm x} \\
t_{\rm y}\\
\end{pmatrix}
\end{split}
\end{equation}
where $t_{\rm x}$ and $t_{\rm y}$ are the component of the translation vector between the two systems, $\alpha$ the rotation angle, $s_{\rm x}$ and $s_{\rm y}$ the scaling factors and $h$ the shear between the transformed axes.

To determine the transformation parameters, we identify the same objects (i.e., nanoparticles) in the two images and estimate their coordinates.
We then perform a nonlinear least-squares fitting of the parameters, minimising the quantity
\begin{equation}
\sum_i \|{\bf r_B}_i-\mathbf{C}({\bf r_A}_i)\|
\end{equation}
where $i$ counts the objects.
Knowing $\mathbf{C}$, the image A can be transformed into the reference system of image B by transforming the coordinate of each pixel in A and interpolating the corresponding intensity to map the position of the pixels in B.

For the case of nanoparticle clusters in FWM, the centroid coordinate position of each nanoparticle cluster from the FWM z-stack was calculated using the "3D object counter" plugin in ImageJ.

{\bf TEM.} Following the FWM analysis, the grids were recovered for TEM analysis by flooding the space between the coverslip and slide with excess water and gently lifting the coverslip\,\cite{HodgsonMMB18}. The grid was subsequently dried and transferred to a 120kV or 200kV transmission EM (Tecnai12 or Tecnai20 respectively, FEI, now Thermo Scientific). The site of interest was retraced using the outline of the sections and calculating the approximate position of the cell(s) of interest. Overview images were collected, followed by subsequent zooms into the area of interest. No fiducials were added, as they are not required in the reported single AuNP probe CLEM. An example of this workflow is described in Fig.\,\ref{S-workflow}.



\section*{Acknowledgements}

This work was funded by the UK EPSRC Research Council (Grants EP/I005072/1, EP/I016260/1, EP/L001470/1, and EP/M028313/1) and the UK BBSRC Research Council (Grants BBL014181/1, BB/M001969/1).

\section*{Authors' contributions}

P. B., W. L. and P.V. conceived the technique and designed the
experiments. W.L. designed the FWM experimental setup and wrote the FWM acquisition software. I.P.
performed all FWM experiments and most of the data analysis. F.M. contributed to the correlation analysis and wrote the corresponding analysis software. L.P. contributed to NP shape analysis and wrote the corresponding analysis software.
P.B. performed part of the analysis and wrote the manuscript.
P.V. performed the internalisation experiments and EM processing. H. T., K. P. A., J. M. and P. V. performed the EM analysis. P.V. contributed to writing parts of the manuscript. All authors discussed and interpreted the results and
commented on the manuscript. 

\section*{Additional information}

{\bf Supplementary Information} is available for this paper.




\end{document}


\title{Supplementary Information - Correlative light electron microscopy using small gold nanoparticles as single probes}


\author[1]{\fnm{Iestyn} \sur{Pope}}

\author[2,4]{\fnm{Hugh} \sur{Tanner}}

\author[1]{\fnm{Francesco} \sur{Masia}}

\author[1]{\fnm{Lukas} \sur{Payne}}

\author[2,5]{\fnm{Kenton Paul} \sur{Arkill}}

\author[2]{\fnm{Judith} \sur{Mantell}}

\author[3]{\fnm{Wolfgang} \sur{Langbein}}

\author*[1]{\fnm{Paola} \sur{Borri}}\email{BorriP@cardiff.ac.uk}

\author*[2]{\fnm{Paul} \sur{Verkade}}\email{P.Verkade@bristol.ac.uk}

\affil[1]{\orgdiv{School of Biosciences}, \orgname{Cardiff University}, \orgaddress{\street{Museum Avenue}, \city{Cardiff}, \postcode{CF10 3AX}, \country{UK}}}

\affil[2]{\orgdiv{School of Biochemistry}, \orgname{University of Bristol}, \orgaddress{\street{University Walk}, \city{Bristol}, \country{UK}}}

\affil[3]{\orgdiv{School of Physics and Astronomy}, \orgname{Cardiff University}, \orgaddress{\street{The Parade}, \city{Cardiff}, \postcode{CF24 3AA}, \country{UK}}}

\affil[4]{Present Address: \orgdiv{Department of Chemistry}, \orgname{Ume\r{a} University}, \city{Ume\r{a}}, \postcode{90187}, \country{Sweden}}

\affil[5]{Present Address: \orgdiv{School of Medicine}, \orgname{University of Nottingham}, \city{Nottingham}, \postcode{NG7 2RD}, \country{UK}}

\maketitle
\thispagestyle{empty}
\pagestyle{empty}

\section{FWM phase versus axial position}\label{phasez}
\begin{figure}[h]%
	\centering
	\includegraphics[width=0.45\textwidth]{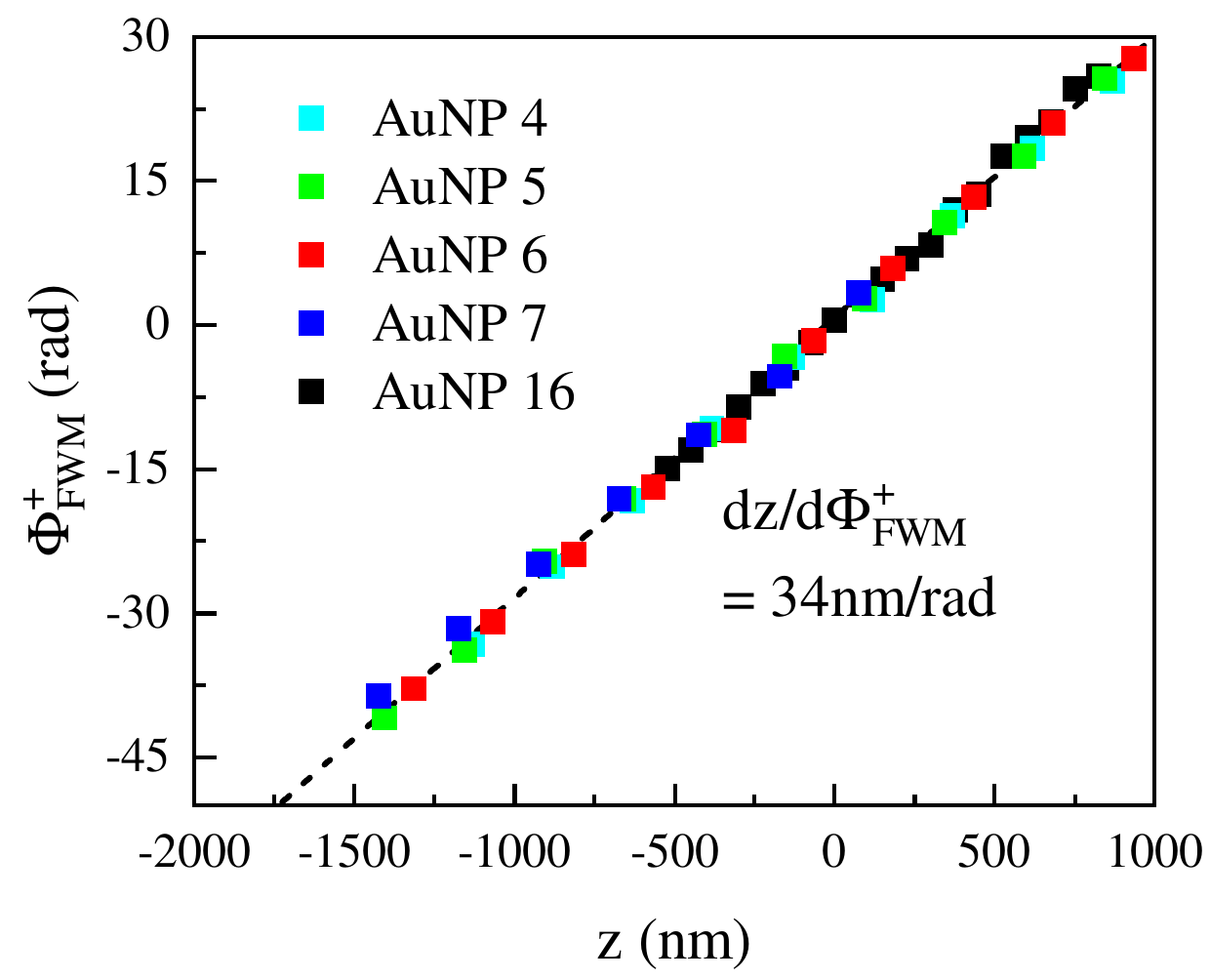}
	\caption{{\bf FWM phase versus NP axial position}. Phase  $\PhiF^+$ of the co-circularly polarised FWM field measured on a set of individual 10\,nm radius AuNPs while scanning their axial position using the sample nano-positioning stage. The phase has been unwrapped by multiples of $2\pi$, and shows a linear dependence on the axial position, as indicated.}\label{zdep}
\end{figure}

We have shown in our previous work that the phase  $\PhiF^+$ of the co-circularly polarised FWM field  encodes the axial displacement between particle and the focus center, thus it can be used to determine the particle $z$ coordinate without axial scanning\,\cite{ZoriniantsPRX17}. This can be easily understood as due to the optical path length difference between the particle and the observation point. For a plane wave of wavevector $k=2{\pi}n/\lambda$ with the refractive
index $n$ of the medium, the phase would be $2kz$, the factor of 2 accounting
for double path in reflection geometry. We have measured $\PhiF^+$ while moving the NP axial position using the sample nano-positioning stage, on a set of 10nm radius AuNPs (see Fig.\ref{zdep}; AuNP 4 to 7 are the same as in Fig.\,\ref{M-CLEM20nm}c while AuNP 16 is additional). We find a linear relationship with a slope $dz/d\Phi=34.2$\,nm/rad, slightly larger than $\lambda/(4\pi
n)=28.8$\,nm/rad. This is due to the propagation of a focussed beam with high NA where a Gouy phase shift occurs, reducing the wavevector
in axial direction due to the wavevector spread in lateral direction. The measured slope is consistent with our previous work\,\cite{ZoriniantsPRX17}.

\section{Analysis of ellipse shapes in TEM}\label{TEMellipse}

\begin{figure*}[h]%
	\centering
	\includegraphics[width=\textwidth]{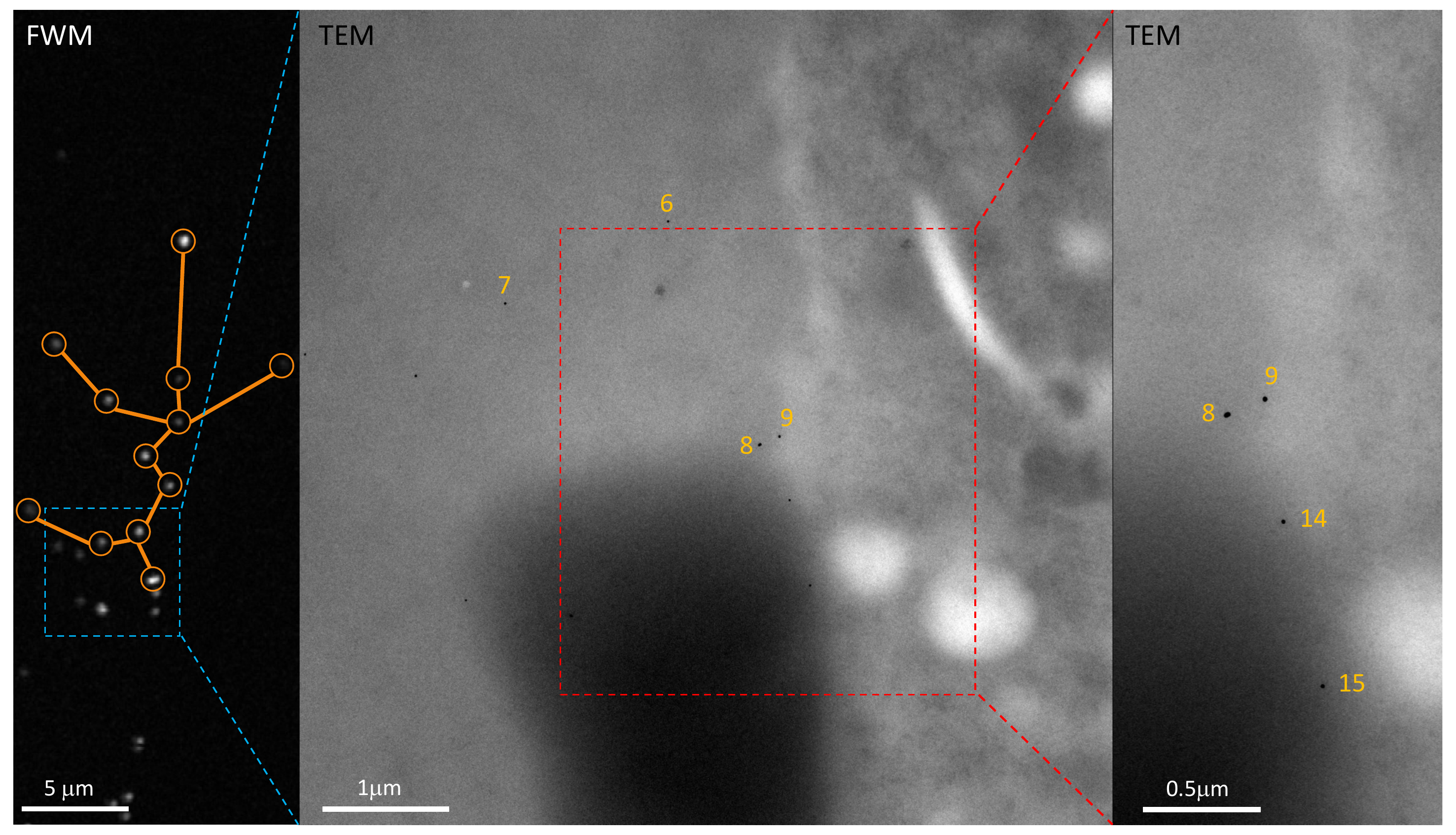}
	\caption{{\bf Additional 10nm radius AuNPs for shape analysis}. Overview TEM compared with the FWM imaging in Fig.\,\ref{M-CLEM20nm}c showing additional AuNPs, numbered as 14 and 15, below those indicated as number 8 and 9. For AuNP 15 a high magnification TEM was acquired and was included in the shape analysis in Fig.\,\ref{M-ellipticity}.}\label{CLEMadd}
\end{figure*}

It was shown in our previous work\,\cite{ZoriniantsPRX17} that using a polarisation-resolved configuration in the FWM field detection provides additional information on the AuNP shape and orientation. Using the CLEM workflow demonstrated here, we have correlatively analysed the measured ratio of the cross to co-circularly polarised FWM field, in amplitude ($\AF^- / \AF^+$) and phase ($\PhiF^- - \PhiF^+$), with the AuNP shape obtained by TEM, and compared the results with the ellipsoid model previously developed\,\cite{ZoriniantsPRX17}. Fig.\,\ref{M-ellipticity} shows the results on a set of AuNPs as labelled in Fig.\,\ref{M-CLEM20nm}c plus an additional AuNP (numbered as 15) for which a high-magnification TEM was also acquired. An overview showing the location of this particle below the AuNP pair 8 and 9 is given in Fig.\,\ref{CLEMadd}.

\begin{figure}[h]%
 	\centering
 	\includegraphics[width=0.45\textwidth]{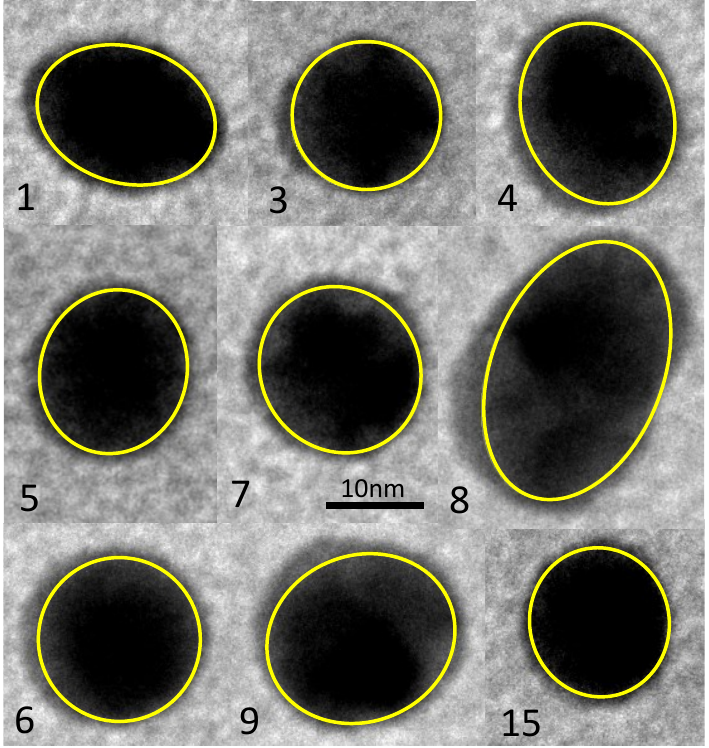}
 	\caption{{\bf Fits of ellipse shape on TEM}.  High-magnification TEM images of selected nanoparticles (numbered as in Fig.\,\ref{M-CLEM20nm}c with the addition of AuNP 15) fitted with an ellipse shape (shown in yellow) using a threshold such that the area of the fitted ellipse is 80\% of the area shown in Fig.\,\ref{M-ellipticity}. }\label{ellipse}
 \end{figure}

As discussed in the Methods section, the fitted ellipses to the TEM images in Fig.\,\ref{M-ellipticity} are obtained using the "Analyse particles - fit ellipse" command in the freely available Java-based image analysis program ImageJ\,\cite{ImageJ}. This command measures and fits objects in thresholded images. It works by scanning the selection until it finds the edge of an object. It then provides the major and minor semi-axis and the orientation angle $\gamma$ of the best fitting ellipse. The orientation angle is calculated between the major axis and a line parallel to the $x$-axis of the image. For the ellipses shown by the yellow lines in the TEM images in Fig.\,\ref{M-ellipticity}, the "auto-threshold" default option was applied. To estimate the error bars in the fitted aspect ratios and in the angle $\gamma$, TEM images were re-fitted using a different threshold such that the area of the fitted ellipse was 80\% of the area obtained with auto-threshold, as shown in Fig.\,\ref{ellipse}. The horizontal errors bars in Fig.\,\ref{M-ellipticity} are the single-sided distances between the values using the auto-threshold option and the re-fitted values.  
 
\section{FWM Ratio}\label{TEMellipse}

\begin{figure*}[b]%
	\centering
	\includegraphics[width=0.9\textwidth]{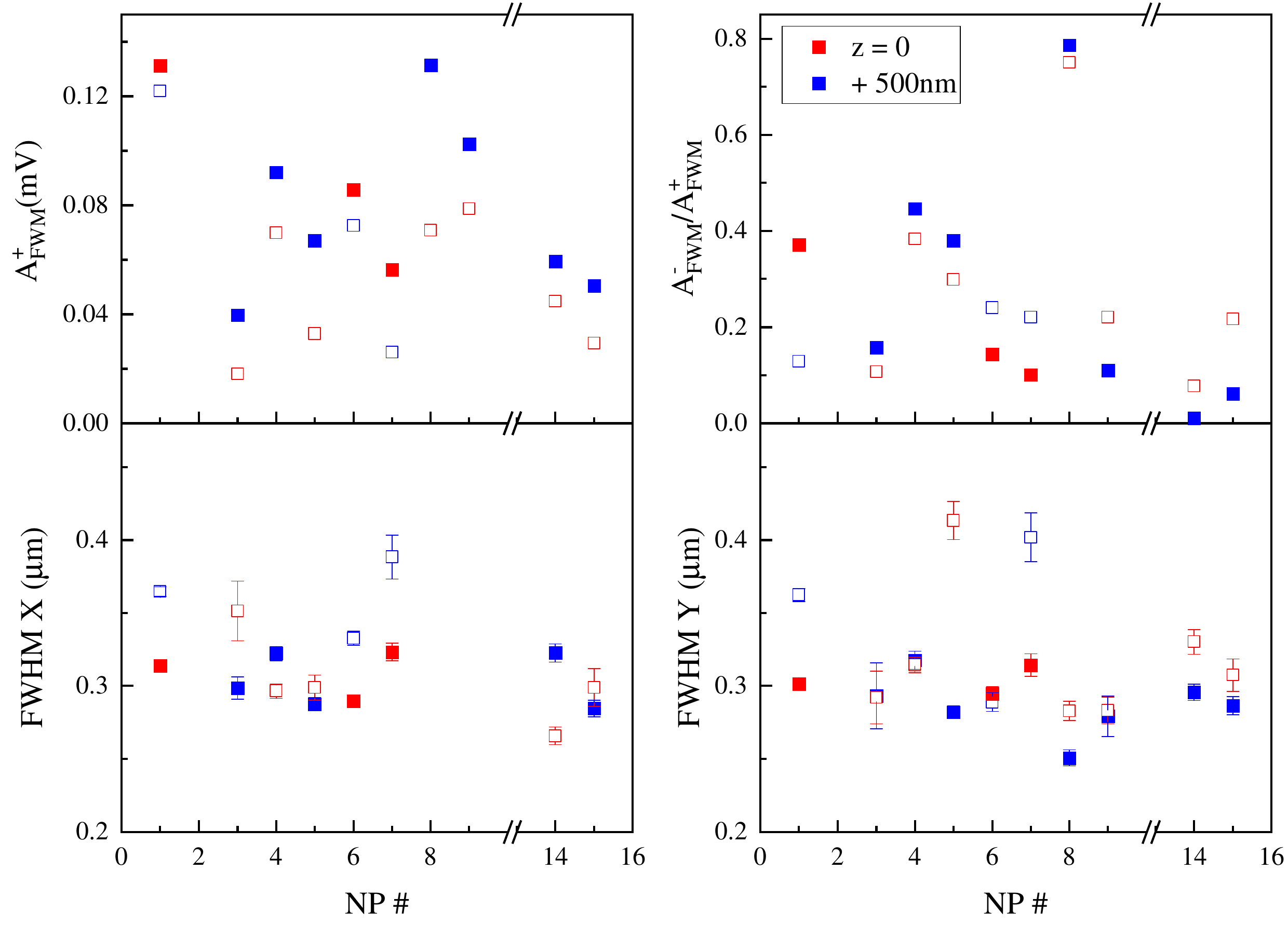}
	\caption{{\bf FWM amplitudes and PSF widths at two focal planes.} Overview of  $\AF^+$, the ratio $\AF^- / \AF^+$, and the full-width at half maximum (FWHM) of the  $\AF^+$ profile in plane along the $x$ and $y$ directions versus NP number, for the AuNPs shown in Fig.\,\ref{M-ellipticity}. Filled (empty) symbols indicate the values at the plane in (out of) focus. The color is coding each plane, as indicated.}\label{ratiovsz}
\end{figure*}

As also discussed in the Methods section, the FWM field ratios in Fig.\,\ref{M-ellipticity} were measured from the two in-plane data sets 0.5\,\textmu m apart in $z$ forming the overview in Fig.\,\ref{M-CLEM20nm}c. However, the FWM ratio values are slightly dependent on the axial position of the AuNP. Hence, care has to be taken to consider the ratio only for NPs that are in focus, based on the maximum co-polarised FWM amplitude detected ($\AF^+$) and on the width of the point-spread function (PSF). An overview of  $\AF^+$, the ratio $\AF^- / \AF^+$, and the full-width at half maximum (FWHM) of the  $\AF^+$ profile in plane along the $x$ and $y$ directions is shown in Fig.\,\ref{ratiovsz} comparing each NP at the two data sets 0.5\,\textmu m apart in $z$. The filled symbols indicate the values at the plane of best focus, used for Fig.\,\ref{M-ellipticity}, while the color is coding each plane, as indicated. For each particle, we see that at the plane of highest amplitude $\AF^+$ the PSF width is small and symmetric in $x$ and $y$, while in the other plane the PSF width increases and in some cases becomes highly asymmetric (see e.g. AuNP 5) due to aberrations. Note that for AuNP 8 and 9 it was not possible to determine a PSF width along $x$, because these two AuNPs are too close to each other, hence only the width in $y$ is given. Note also that AuNP 14 was excluded from the analysis in Fig.\,\ref{M-ellipticity} because its PSF width was found to be significantly asymmetric in both planes.   

\subsection{Model of FWM ratio versus AuNP ellipticity}\label{TEMellipse_model}
To explain the model, we start with a description of the polarizability of the AuNP\,\cite{PayneNS20}. In its own reference frame, we choose for the semi-axes ($a$, $b$, $c$) of an ellipsoidal AuNP to be aligned, respectively, along the orthogonal axes ($x'$, $y'$, $z'$) of a Cartesian coordinate system in the positive directions. The polarizability tensor describing the AuNP in its own reference frame is then given by

\be\label{eq:polten}
\hpol'=\begin{pmatrix}
	\alpha_a & 0 & 0\\
	0 & \alpha_b & 0\\
	0 & 0 & \alpha_c
\end{pmatrix}.
\ee
The components of $\hpol'$, ($\alpha_\mathrm{a}$, $\alpha_\mathrm{b}$, $\alpha_\mathrm{c}$) are related to the unmodified (pump-modified) relative complex permittivity of the AuNP, \eu\ (\ep), and the relative permittivity of the surrounding medium, \ed, with

\be\label{eq:alcomp}
\aj = V\ez\frac{\epsilon-\ed}{\ed+\Lj(\epsilon-\ed)},
\ee
where, $V=\frac{4}{3}\,\pi\,a b c$, is the volume of the AuNP, \Lj\ are the factors describing the geometry of the ellipsoidal shape, $\epsilon$ has the value of either \eu\ for a particle in the absence of the pump, or \ep\ in presence of the pump pulse at 0.5\,ps delay, and \ez\ is the permittivity of free space. We assume \ed\ to be constant as a function of the wavelength, $\lambda$, of the incident radiation. \Lj\ are found from the analytical expressions,

\be\label{eq:Lj}
\Lj = \frac{abc}{2}\int_0^\infty\frac{\mathrm{d}q}{(j^2+q)\sqrt{(a^2+q)(b^2+q)(c^2+q)}}
\ee

The AuNP polarizability is transformed into the laboratory reference frame, whose axes we label ($x$, $y$, $z$), with

\be\label{eq:rot}
\hpol = \hrot\hpol'\hrot^\mathrm{T},
\ee
We define \hrot\ and $\hrot^\mathrm{T}$ as

\be\label{eq:rotmax}
\hrot = \Rpsi\Rthe\Rphi,\qquad\hrot^\mathrm{T} = \Rphi^\mathrm{T}\Rthe^\mathrm{T}\Rpsi^\mathrm{T}
\ee
with

\be\label{eq:Rphi}
\Rphi=\begin{pmatrix}
	1 & 0 & 0\\
	0 & \cos(\phi) & -\sin(\phi)\\
	0 & \sin(\phi) & \cos(\phi)
\end{pmatrix},
\ee

\be\label{eq:Rthe}
\Rthe=\begin{pmatrix}
	\cos(\theta) & 0 & \sin(\theta)\\
	0 & 1 & 0\\
	-\sin(\theta) & 0 & \cos(\theta)
\end{pmatrix},
\ee

\be\label{eq:Rpsi}
\Rpsi=\begin{pmatrix}
	\cos(\psi) & -\sin(\psi) & 0\\
	\sin(\psi) & \cos(\psi) & \\
	0 & 0 & 1
\end{pmatrix}.
\ee

The AuNP permittivity is modelled with the function\,\cite{MasiaPRB12}

\be\label{eq:perm}
\eu = 1 - \frac{\op^2}{\omega(\omega+i\Gamma)}+\eb(\omega)
\ee
where $\omega=2\pi c/\lambda_0$, $c$ is the speed of light in vacuum, $\Gamma$ is the relaxation rate, \eb\ is the contribution due to bound electrons associated with interband transition from the $d$ bands to the conduction band, and $\op=n_\mathrm{e}e^2/\ez m_0$ is the plasma frequency ($n_\mathrm{e}$, $e$, $m_0$ are the conduction electron density, charge, and effective mass, respectively).

The permittivity as function of the optical excitation is modelled as described in Ref.\cite{MasiaPRB12}. We call the value without excitation $\epsilon_{\rm u}$, and determine the value $\epsilon_p$ with excitation by a pulse at 550\,nm wavelength, 0.1pJ/\textmu m$^2$ fluence and 0.5\,ps delay.

The polarization of the AuNP is given by
\be\label{eq:polv}
\mathbf{p} = \hpol\mathbf{E}
\ee
where $\mathbf{E}$ is the incident field and the bold font indicates that it is a vector quantity. The pump field induces a change in the polarizability given by

\be\label{eq:delp}
\Delta\hpol = \hpol_{\rm u}-\hp,
\ee

using $\epsilon_{\rm u}$ and $\epsilon_{\rm p}$, respectively.
The FWM field resulting from the subsequent probe of the AuNP is proportional to the modified polarizability and the probe field, so that in suited units we can write $\efwm=\Delta\hpol\epro$, where \epro\ is the probe field.

We note that \epro\ has circular polarisation at the sample, and choose to compactly represent it here using the Jones vector 

\be\label{eq:probe}
\epro=\begin{pmatrix}
	1\\
	i\\
	0
\end{pmatrix}.
\ee
We define the co- and cross-polarised components of \efwm\ as

\be\label{eq:eco}
\eco = \epro^\star\,\cdot\,\Delta\hpol\epro
\ee
and

\be\label{eq:ecro}
\ecro = \epro\,\cdot\,\Delta\hpol\epro,
\ee
respectively, where ($\star$) indicates the complex conjugate.

Hence, we can determine the amplitude and phase ratios of the co- and cross-polarized projections of the FWM field from

\be\label{eq:amp}
\frac{\afm}{\afp} = \left\vert\frac{\ecro}{\eco}\right\vert
\ee
and

\be\label{eq:phase}
\pfm -\pfp = \arg\left(\frac{\ecro}{\eco}\right)\,.
\ee

For the plot in Fig.\,\ref{M-ellipticity}, when the NP $a$ and $c$ axis are tilted by 45 degrees in the $x,z$ plane, the projected minor to major axis ratio in the $x$-direction is given by $\sqrt{2}/\sqrt{(a/b)^2+1}$ for the prolate case ($a>b$) and by $\sqrt{(a/b)^2+1}/\sqrt{2}$ for the oblate case ($a<b$).

\section{FWM-EM correlation accuracy}\label{CLEMaccuracy}

\begin{figure*}[b]%
	\centering
	\includegraphics[width=0.6\textwidth]{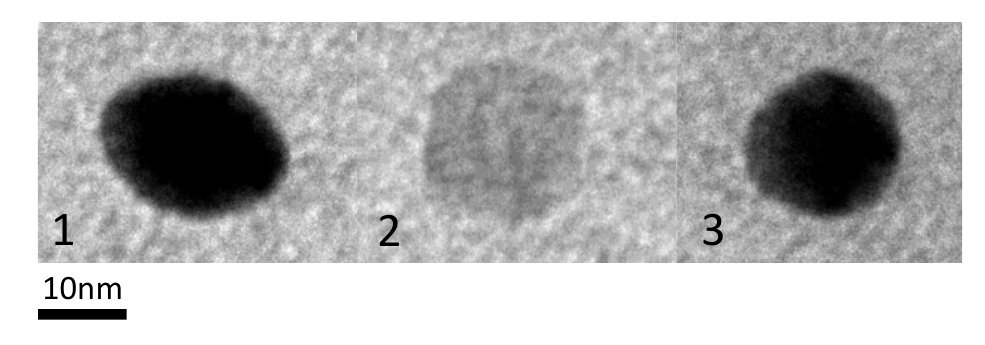}
	\caption{{\bf TEM of AuNP n.2.} High magnification TEM showing AuNP 2 compared with 1 and 3.}\label{TEMAuNP2}
\end{figure*}

\begin{figure*}[b]%
	\centering
	\includegraphics[width=\textwidth]{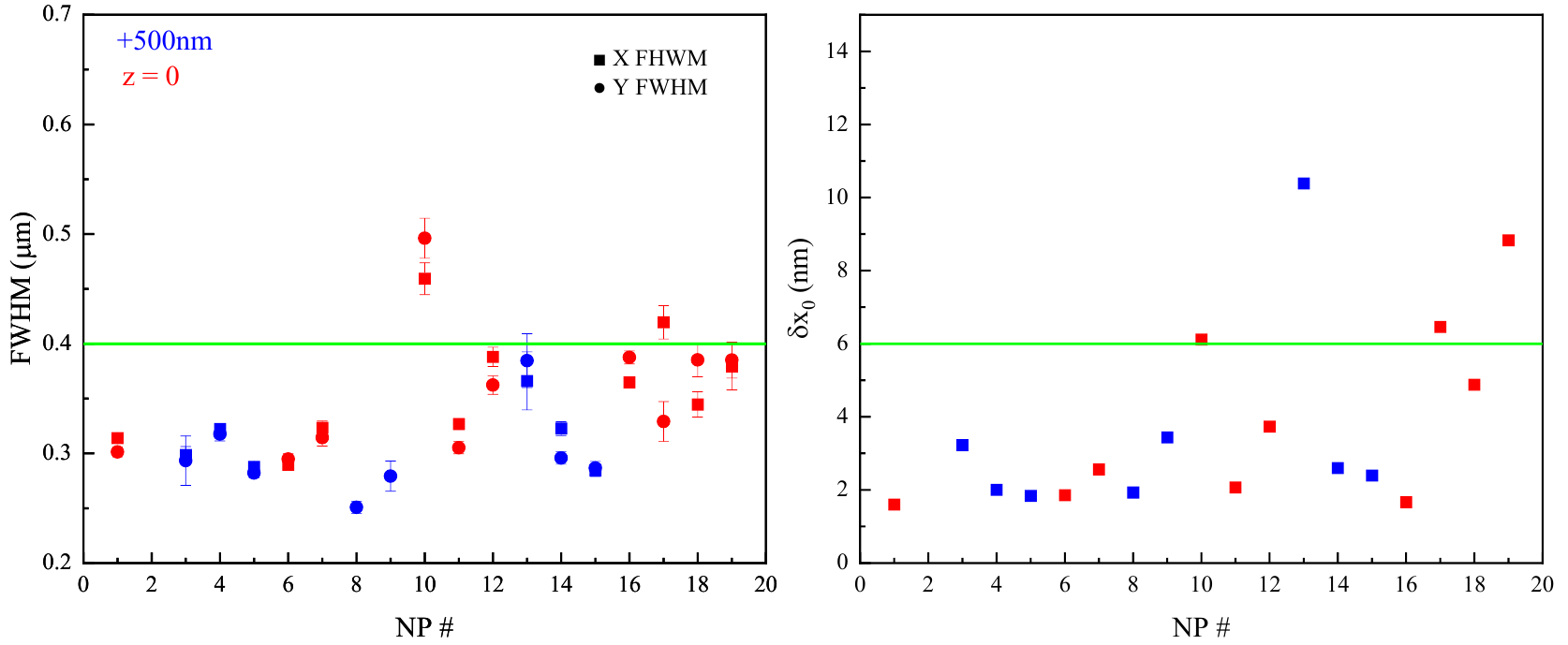}
	\caption{{\bf 10-nm radius AuNP selection for CLEM}  Left: Overview of the FWHM of the  $\AF^+$ profile in plane crossing the AuNP centre along the $x$ and $y$ directions versus NP number, for the AuNPs shown in Fig.\,\ref{M-CLEM20nm} and Fig.\,\ref{20nm_correlation_ED}. Right: Error in the centroid localisation from Gaussian fits of $\AF^+$ line-profiles crossing the AuNP centre along the $x$ direction. Values are shown at the plane of highest $\AF^+$ amplitude, and the colour is coding each plane, as indicated. Lines show the cut-off criteria for AuNP selection (see text), which resulted in excluding AuNP 10, 13, 17, 19.}\label{PSF20nm}
\end{figure*}

\begin{figure*}[h]%
	\centering
	\includegraphics[width=1\textwidth]{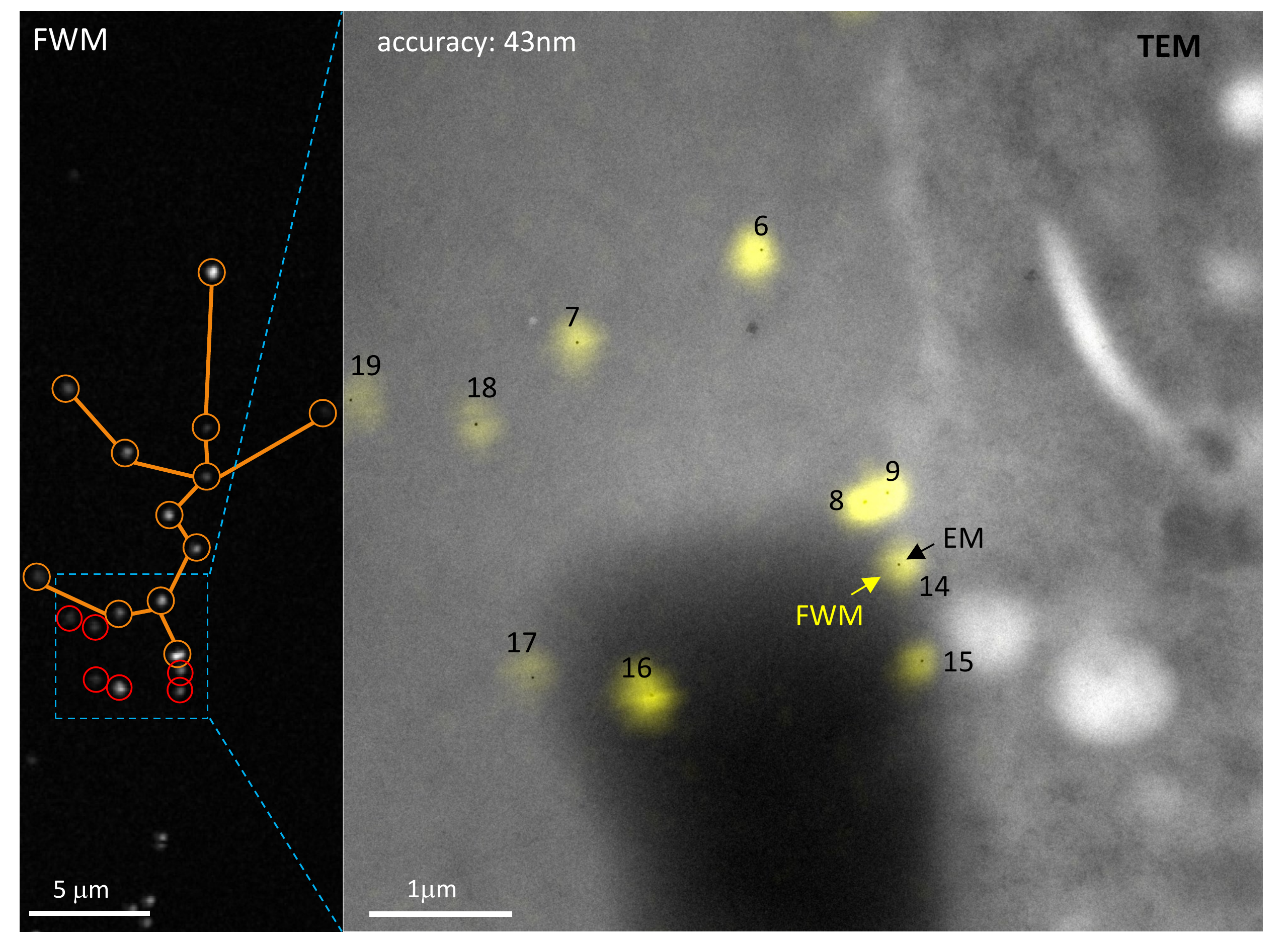}
	\caption{{\bf CLEM correlation accuracy}. Overlay of FWM field amplitude and TEM image using a different region compared to Fig.\,\ref{M-CLEM20nm}c as indicated by the blue frame on the left side (red circles highlight the additional AuNPs). The FWM image is transformed into the EM reference system using a linear transformation matrix that accounts for translation, rotation, shear and scaling of axes. On the right side, yellow spots show individual AuNPs in FWM overlaid onto the EM (black dots). The correlation accuracy is indicated. AuNPs 17 and 19 were out of focus and excluded from the correlation analysis (see text).}\label{20nm_correlation_ED}
\end{figure*}

As stated in the paper, we evaluate the correlation accuracy using the coordinates of AuNPs measured in FWM and in TEM, and transforming the coordinates of AuNPs from FWM into the reference system of the TEM image, using a linear transformation. Apart from AuNP 2 which was excluded due to its low FWM amplitude and atypical TEM contrast (see Fig.\,\ref{TEMAuNP2}), we applied selection criteria to exclude AuNPs which are too out of focus in FWM by performing the following analysis. For all AuNPs in the dataset, we considered the FWHM of the $\AF^+$ profile in plane crossing the AuNP centre along the $x$ and $y$ and the error in the centroid localisation from Gaussian fits of the line-profiles. This is shown in Fig.\,\ref{PSF20nm} for the 10\,nm-radius AuNPs in Fig.\,\ref{M-CLEM20nm} and Fig.\,\ref{20nm_correlation_ED}. We excluded AuNPs having a FWHM larger than 0.4\,\textmu m as deemed too out of focus, hence subject to localisation uncertainties originating from aberrations in the microscope objective as well as deformations of the pioloform layer supporting the resin section which change from FWM in water to TEM in vacuum. These excluded AuNP 10 and 17, for which the centroid localisation error was found to be around 6\,nm, more than twice the average value observed for AuNPs in focus. By inspecting the localisation error, we then also excluded AuNP 13 and 19, which had a centroid localisation error around 10\,nm (and a FWHM near the 0.4\,\textmu m cut-off), to ensure that the localisation precision is not the limiting factor in the correlation analysis.

Since the resin section is only 300\,nm thick, smaller than the axial extension of the PSF in FWM imaging (as shown in Fig.\,\ref{M-CLEM20nm}b) one would not expect to have AuNPs out of focus. On the other hand, we observed that the pioloform layer supporting the resin section during FWM imaging was not flat but rather exhibited bending and wrinkles. Indeed, by exploiting the topography information encoded in the phase of the reflected probe field, we reconstructed a height profile for the region imaged in Fig.\,\ref{M-CLEM20nm} and Fig.\,\ref{20nm_correlation_ED}. This was achieved by unwrapping the phase profile (i.e. removing $2\pi$ phase jumps to have a continuous phase change) and using the relationship between phase and axial position from the slope $dz/d\Phi=34$\,nm/rad (see also Fig.\ref{zdep}). The resulting height profile is shown in Fig.\,\ref{M-surface} and illustrates that AuNP 10,13,17, and 19 are indeed positioned at a significantly different height than the others, consistent with them appearing out of focus compared to other AuNPs in the image.  

\begin{figure*}[b]%
	\centering
	\includegraphics[width=\textwidth]{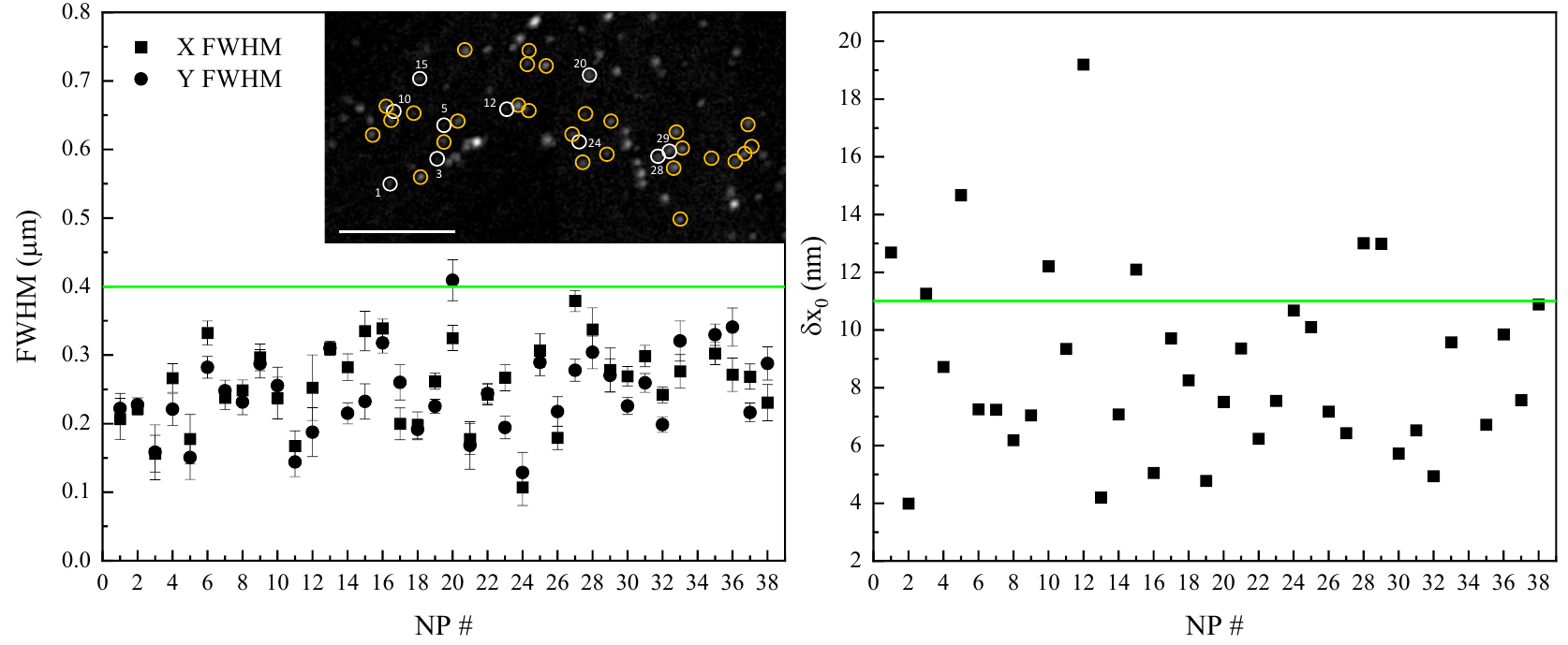}
	\caption{{\bf 5-nm radius AuNP selection for CLEM}  Left: Overview of the FWHM of the  $\AF^+$ profile in plane crossing the AuNP centre along the $x$ and $y$ directions versus NP number, for the AuNPs shown in Fig.\,\ref{M-10nm_correlation_ED} and Fig.\,\ref{CLEM10nm}. Right: Error in the centroid localisation from Gaussian fits of $\AF^+$ line-profiles crossing the AuNP centre along the $x$ direction. Lines show the cut-off criteria for AuNP selection (see text), which resulted in excluding  6 AuNPs (number 1, 3, 5, 10, 12, 15) for the data shown in Fig.\,\ref{M-10nm_correlation_ED} and 4 AuNPs (number 20, 24, 28, 29) for the data in Fig.\,\ref{CLEM10nm}. The inset highlights the excluded AuNPs in the FWM $\AF^+$ image as white circles (with corresponding numbers) and the included AuNPs as yellow circles. Scale bar in the inset image is 5\,\textmu m.}\label{PSF10nm}
\end{figure*}

Selection criteria applied to the 5\,nm-radius AuNPs are shown in Fig.\,\ref{PSF10nm}. Also here,  we considered the FWHM of the $\AF^+$ profile in plane crossing the AuNP centre along the $x$ and $y$ and the error in the centroid localisation from Gaussian fits of the line-profiles. As for the 10\,nm-radius AuNP selection criteria, particles having a FWHM larger than 0.4\,\textmu m were excluded, which led to excluding AuNP 20. By inspecting the localisation error, we see that for all particles this is larger (at least twice) than what observed for the 10\,nm-radius AuNPs in Fig.\,\ref{PSF20nm}, as expected considering the lower signal to noise ratio from the scaling of the FWM field amplitude with the AuNP size. We therefore applied a higher cut-off to this dataset, and excluded AuNPs having a centroid localisation error larger than 11\,nm, to retain the majority of 5\,nm-radius AuNPs, while the cut-off is still significantly below, and thus not limiting, the final accuracy found. 

\begin{figure*}[b]%
	\centering
	\includegraphics[width=\textwidth]{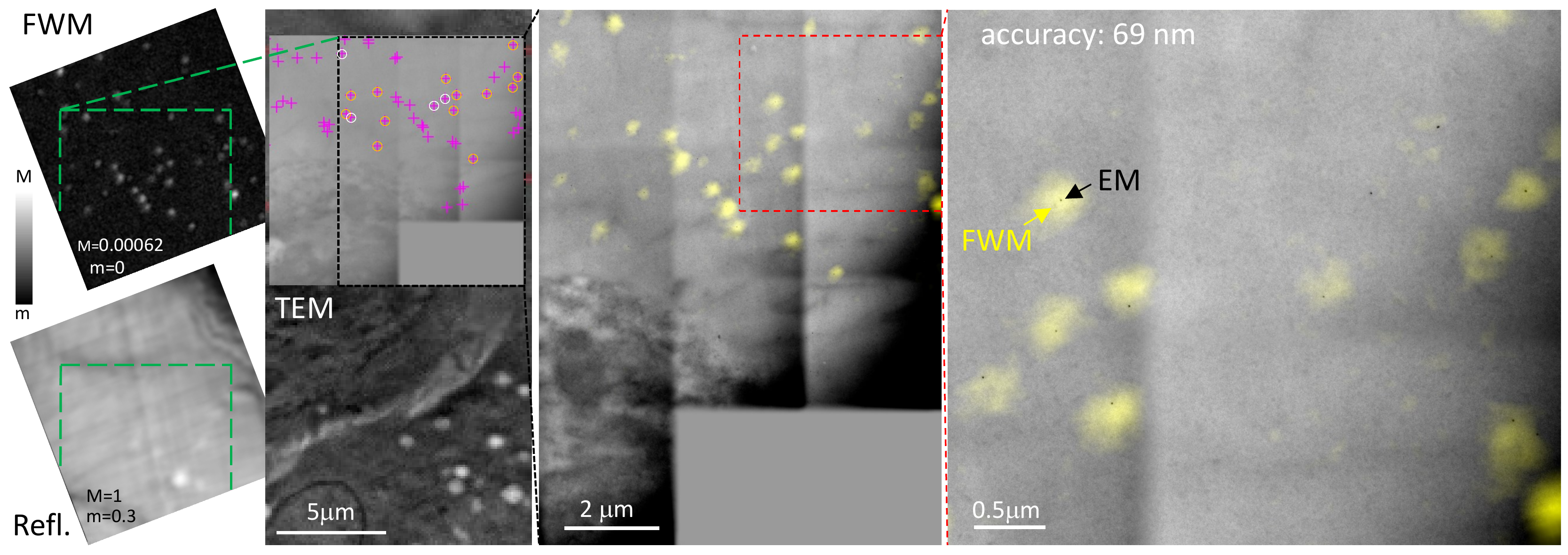}
	\caption{{\bf CLEM correlation accuracy with 5\,nm-radius AuNPs}. HeLa cells incubated with 5nm-radius AuNPs bound to the EGF protein. CLEM overview on a region adjacent to the one shown in Fig.\,\ref{M-10nm_correlation_ED}. Individual AuNPs are detected background-free in FWM (left). The confocal reflection image simultaneously acquired with FWM is shown below (linear grey scales are from m to M as indicated; M=1 corresponds to 65\,mV rms detected, see Methods for details of the excitation and detection conditions). A large overview TEM of the same region is shown together with a series of high resolution EM images stitched together. Individual AuNPs are highlighted by the circles (see dashed green frame for the corresponding AuNPs in FWM). The overlay between FWM (yellow) and TEM (grey) is shown on the center and further zoomed into the indicated red dashed area on the right side (contrast adjusted to aid visualisation). For the correlation analysis, of the 17 individual AuNPs highlighted by the circles, 4 (white circles) were discarded as being of focus. The FWM image was transformed into the EM reference system using a linear transformation matrix that accounts for translation, rotation, shear and scaling of axes. On the right side, individual AuNPs identified in FWM (yellow spots) are seen in EM (black dots). The correlation accuracy is indicated.}\label{CLEM10nm}
\end{figure*}

 An example of CLEM with FWM imaging using 5\,nm-radius AuNPs in HeLa cells is shown in Fig.\,\ref{M-10nm_correlation_ED}. Several AuNPs are clearly visible in both FWM and TEM. A few AuNPs are too close to be spatially distinguished in the FWM image, but 19 individual AuNPs are available for position analysis. This resulted in a correlation accuracy of 58\,nm, retaining 13 AuNPs for the correlation (see orange circles in Fig.\,\ref{M-10nm_correlation_ED}), while 6 AuNPs were excluded (white circles in Fig.\ref{M-10nm_correlation_ED}) based on the criteria discussed above.  Another example showing an adjacent region is provided in Fig.\,\ref{CLEM10nm} where again 13 individual AuNPs were used for the correlation. Merging both regions results in a correlation accuracy of 63\,nm.
 
\begin{figure*}[h]%
	\centering
	\includegraphics[width=1\textwidth]{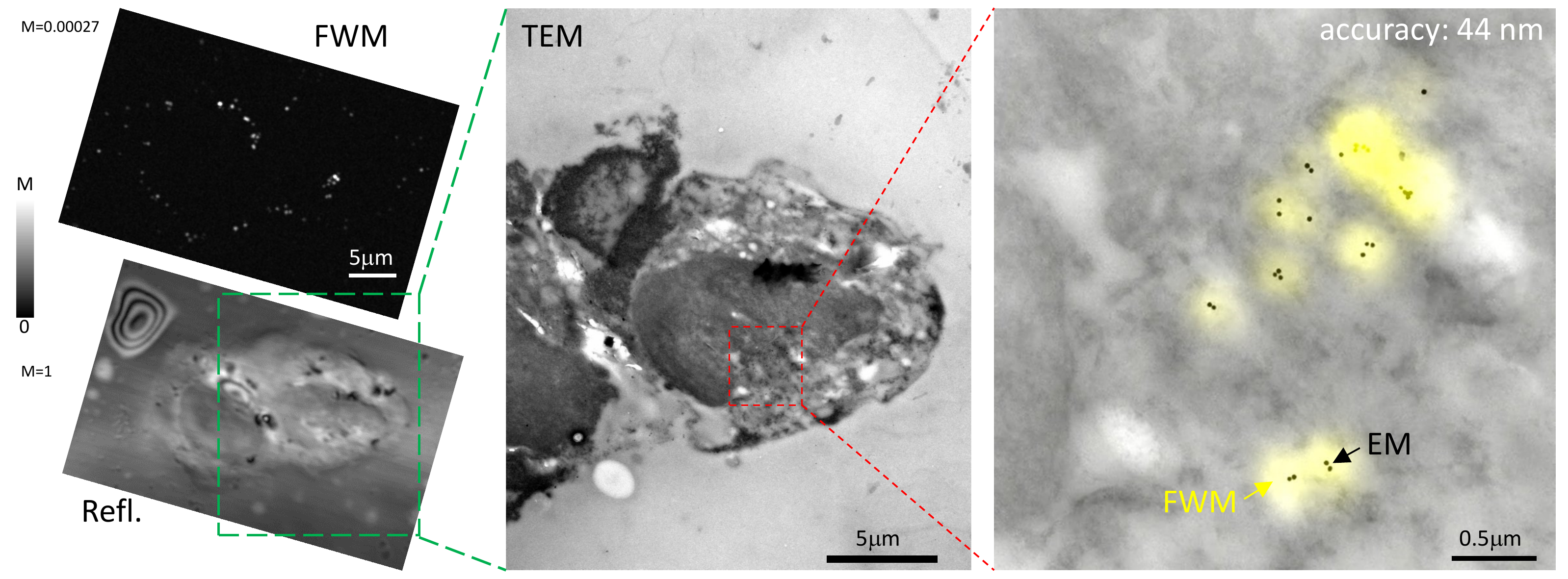}
	\caption{{\bf CLEM correlation accuracy with 3D FWM analysis.} FWM-CLEM using 10nm-radius AuNPs bound to EGF internalised in HeLa cells whereby the coordinates of the particles in FWM are measured in 3D via a z-stack. A large FWM overview in 2D with corresponding reflection image measured simultaneously is shown on the left (linear grey scales are from 0 to M as indicated; M=1 corresponds to 33mV rms detected; see Methods for details of the excitation and detection conditions). A TEM overview of the same region is shown in the center, as indicated by the green dashed frame. On the left, an overlay of FWM field amplitude (yellow) and TEM image (grey) is shown for the region highlighted by the red dashed frame, where FWM is a maximum amplitude projection from a 3D z-stack (50\,nm step size in $z$). AuNPs form small clusters and are no longer resolved as individual particles in FWM. The centroid position of each cluster was determined in 3D from the FWM z-stack (see Methods), and its 2D in-plane coordinates were compared with the position of the geometrical centre of the cluster in TEM (which is a 2D transmission projection) for the correlation analysis. The resulting correlation accuracy from the comparison of the ten clusters shown in the figure is indicated.}\label{SM20nm_cluster}
\end{figure*}

 Considering that the shot-noise limited precision in locating the centroid position of a AuNP in focus by FWM is only a few nanometres, the measured correlation accuracy is limited by systematic errors. To address this point, we performed FWM-CLEM using 10\,nm-radius AuNPs  whereby the coordinates of the particles in FWM were measured in 3D with a fine axial scan (50\,nm step size in $z$), such that the coordinates at the plane of optimum focus are accurately determined and systematics from e.g. out-of-focus aberrations are eliminated. These results are summarized in Fig.\,\ref{M-20nm_cluster}. AuNPs form small clusters and are no longer resolved as individual particles in FWM. Therefore, in this case, we determined the centroid position of the cluster in 3D from the FWM z-stack (see Methods), and compared its 2D in-plane coordinates with the position of the geometrical centre of the cluster in TEM (which is a 2D transmission projection) for the correlation analysis. The resulting correlation accuracy for the six clusters shown in Fig.\,\ref{M-20nm_cluster} is 36\,nm. Another example correlating 10 clusters is provided in Fig.\,\ref{SM20nm_cluster}, for which an accuracy of 44\,nm is found. 
 
\begin{figure*}[b]%
	\centering
	\includegraphics[width=\textwidth]{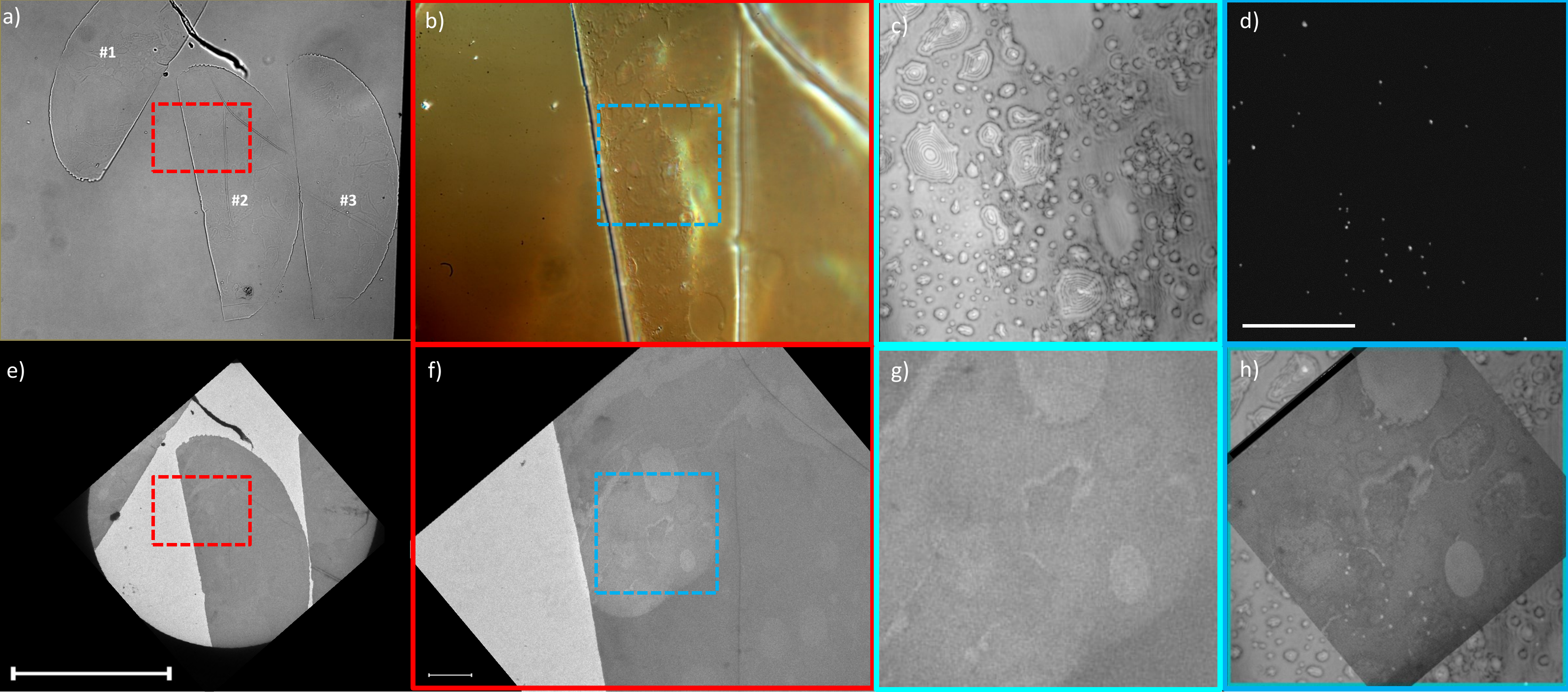}
	\caption{{\bf CLEM workflow.} Workflow of FWM-CLEM acquisition. Sections are first visualised by bright-field transmission and DIC microscopy available in the FWM microscope, followed by simultaneous reflection and FWM imaging. Sections are then retrieved for TEM analysis, where the same cell regions are identified. a) Brightfield transmission microscopy overview image (using a $10\times$ 0.3NA dry objective with $1\times$ tube lens). It shows 3 sections with some folds in section 2 and another imperfection at the top of section 2. b) Higher magnification DIC image (red boxed area from a), using a $60\times$ 1.27 NA water objective, $1\times$ tube lens 1.34 NA oil condenser; the folds are visible as well as the outlines of cells and other features in the section. c) Confocal reflection of the corresponding blue boxed area (on an amplitude log scale). d) FWM acquired simultaneously with reflection, as a maximum intensity projection over a z-stack (on an amplitude log scale, contrast adjusted for visual purposes). e) The grid is retrieved for TEM analysis and a similar overview image is acquired. f) Higher magnification TEM showing the same folds and outlines as in b, as highlighted by the blue frame. g) Magnified crop of the TEM area in f where parts of the cells are recognised as seen in confocal reflection c. h) Overlay of the reflection, FWM and TEM area. FWM was acquired with a pump-probe delay time of 0.5 ps, pump (probe) power at the sample of 30\,\textmu W (15\,\textmu W), 1\,ms-pixel dwell time, pixel size in plane 72\,nm, 500\,nm step size in $z$ and 13 z-steps (6\,\textmu m total range). Scale bar is 250\,\textmu m in a and d, 20\,\textmu m in b and f, and 20\,\textmu m in c,d,g, and h.}\label{workflow}
\end{figure*}

 An example of CLEM workflow is shown in Fig.\,\ref{workflow}. As a first step, a brightfield light transmission microscopy overview image is acquired (using a $10\times$, 0.3NA dry objective with $1\times$ tube lens), see Fig.\,\ref{workflow}a. It shows 3 sections with some folds in section 2 and another imperfection at the top of section 2. Fig.\,\ref{workflow}b shows a higher magnification differential interference contrast (DIC) image (red boxed area from Fig.\,\ref{workflow}a, using a $60\times$ 1.27 NA water objective, $1\times$ tube lens 1.34 NA oil condenser) where the folds are visible as well as the outlines of cells and other features in the section. Fig.\,\ref{workflow}c shows the confocal reflection of the corresponding blue boxed area, and Fig.\,\ref{workflow}d the FWM acquired simultaneously with reflection, as a maximum intensity projection over a z-stack (on an amplitude log scale, contrast adjusted for visual purposes). FWM was acquired with a pump-probe delay time of 0.5 ps, pump (probe) power at the sample of 30\,\textmu W (15\,\textmu W), 1\,ms-pixel dwell time, pixel size in plane 72\,nm, 500\,nm step size in $z$ and 13 z-steps (6\,\textmu m total range). In Fig.\,\ref{workflow}e, the grid is retrieved for TEM analysis and a similar overview image is acquired (compare Fig.\,\ref{workflow}a). Fig.\,\ref{workflow}f shows the same folds and outlines as in Fig.\,\ref{workflow}b, as highlighted by the blue frame. Fig.\,\ref{workflow}g shows a magnified crop of the TEM area in Fig.\,\ref{workflow}f where parts of the cells are recognised as seen in confocal reflection Fig.\,\ref{workflow}c. Fig.\,\ref{workflow}h shows an overlay of the reflection, FWM and TEM area.

